\documentclass[]{tJOT2e_modif}

\usepackage{bm}
\usepackage{graphicx}

\begin{document}
%\doi{10.1080/14685248.YYYYxxxxxx}
% \issn{1468-5248}
% \jvol{00} \jnum{00} \jyear{2011}
\articletype{}

\newcommand{\bR}{\bm R}
\newcommand{\R}{R}
\newcommand{\bV}{\bm V}
\newcommand{\bA}{\bm A}
\newcommand{\Rey}{R_\lambda}
\newcommand{\lagav}[1]{\left\langle #1 \right\rangle_{\R_0}}

\title{Geometry and violent events in turbulent pair dispersion}

\author{Rehab Bitane, Holger Homann, and J\'er\'emie
  Bec$^{\ast}$\thanks{$^\ast$Corresponding author. Email:
    jeremie.bec@oca.eu \vspace{6pt}}\\\vspace{6pt} {\em{Laboratoire
      J.-L.\ Lagrange UMR 7293, Universit\'e de Nice-Sophia Antipolis,
      CNRS, Observatoire de la C\^ote d'Azur, BP 4229, 06304 Nice
      Cedex 4, France}}\\\vspace{6pt}\received{\today} }

\maketitle

\begin{abstract}
  The statistics of Lagrangian pair dispersion in a homogeneous
  isotropic flow is investigated by means of direct numerical
  simulations. The focus is on deviations from Richardson
  eddy-diffusivity model and in particular on the strong fluctuations
  experienced by tracers. Evidence is obtained that the distribution
  of distances attains an almost self-similar regime characterized by
  a very weak intermittency. The timescale of convergence to this
  behavior is found to be given by the kinetic energy dissipation time
  measured at the scale of the initial separation. Conversely the
  velocity differences between tracers are displaying a strongly
  anomalous behavior whose scaling properties are very close to that
  of Lagrangian structure functions. These violent fluctuations are
  interpreted geometrically and are shown to be responsible for a
  long-term memory of the initial separation. Despite this strong
  intermittency, it is found that the mixed moment defined by the
  ratio between the cube of the longitudinal velocity difference and
  the distance attains a statistically stationary regime on very short
  timescales. These results are brought together to address the
  question of violent events in the distribution of distances. It is
  found that distances much larger than the average are reached by
  pairs that have always separated faster since the initial time. They
  contribute a stretched exponential behavior in the tail of the
  inter-tracer distance probability distribution. The tail approaches
  a pure exponential at large times, contradicting Richardson
  diffusive approach. At the same time, the distance distribution
  displays a time-dependent power-law behavior at very small values,
  which is interpreted in terms of fractal geometry. It is argued and
  demonstrated numerically that the exponent converges to one at large
  time, again in conflict with Richardson's distribution.

\end{abstract}

\section{Introduction}
\label{sec:intro}
It is known since Taylor's seminal work \cite{taylor:1921} that
tracers transported by a turbulent flow approach a diffuse behavior on
time scales much longer than the Lagrangian correlation time of the
flow.  These ideas are now commonly used in applications, as for
instance in air quality control, to model effective mixing properties
in terms of an eddy diffusivity. Such models give a good handle on
long-term averages and are successfully used to determine, for
instance, possible health hazards linked to a long exposure downstream
a pollutant source.  However they are unable to capture strong local
fluctuations stemming from the complex structure of the turbulent
flow. Such events cannot be directly predicted from the averaged
concentration field as they relate to higher-order moments. Accessing
these fluctuations is crucial in order to quantify for instance the
likeliness of finding a local concentration exceeding a high
threshold.

Second-order statistics, such as the variance of a transported
concentration field and more generally the spatial correlation of a
passive scalar, are statistically related to the relative motion of
tracers (see, e.g., \cite{monin-yaglom:1971}). The problem consists
then in investigating the time evolution of the separation $\bR(t) =
\bm X_1(t) - \bm X_2(t)$ between two Lagrangian trajectories. In
turbulence, the distance $|\bR|$ follows Richardson--Obukhov
superdiffusive law $\langle |\bR(t)|^2 \rangle \sim \epsilon\,t^3$,
where $\epsilon$ is the mean rate of kinetic energy dissipation. The
long-term behavior is thus becoming independent of the initial
separation $|\bR(0)|=r_0$, whence the designation of \emph{explosive}
pair separation. In his original arguments to derive this law,
Richardson \cite{richardson:1926} assumed that pair separation is a
diffusion process with a scale dependent diffusivity $K$. Then the
transition probability density $p(r,t\,|\,r_0,0)$ that two tracers are
at a distance $|\bR(t)|=r$ knowing that they were initially separated
by $|\bR(0)|=r_0$ satisfies the Fokker--Planck equation
\begin{equation}
  \partial_t p = \frac{1}{r^2}\partial_r\left[ r^2
    K(r) \,\partial_r p\right].
  \label{eq:diffusion}
\end{equation}
The atmospheric measurements of Richardson, later refined by
Obukhov~\cite{obukhov:1941} in view of Kolmogorov 1941 phenomenology,
led to assume that for separations $r$ within the inertial range of
turbulence $K(r)\propto\epsilon^{1/3}r^{4/3}$. This implies that, at
large times, $\langle |\bR(t)|^2\rangle \propto\epsilon\,t^3$ and the
transition probability reads
\begin{equation}
  p(r,t\,|\,r_0,0) \propto \frac{r^2}{\langle |\bR(t)|^2\rangle^{3/2}}
  \,\,\exp\left[-\frac{A\,r^{2/3}}{\langle |\bR(t)|^2
      \rangle^{1/3}}\right],
  \label{eq:pdf_richardson}
\end{equation}
where $A$ is a positive constant. The considerations leading to the
Fokker--Planck equation (\ref{eq:diffusion}) rely on the use of a
central-limit theorem for $\bR(t)$ and thus on the hypothesis that the
velocity difference between the two tracers is correlated over
timescales much smaller than those of interest.  As noticed for
instance in \cite{falkovich-etal:2001}, such an hypothesis can hardly
be invoked. It is indeed known in turbulence that eddies of size $r$
are correlated over a time of the order of their turnover time $\tau_r
\sim \epsilon^{-1/3}r^{2/3}$. Hence for separations that grow like $r
\sim \epsilon^{1/2}t^{3/2}$, one has $\tau_r \sim t$ , so that the
dominant flow structures in the separation dynamics are in principle
correlated over timescales of the order of the observation
time. Despite such shortcomings, the diffusive approach proposed by
Richardson and in particular the explicit form
(\ref{eq:pdf_richardson}) for the transition probability density have
proven being relevant in some
asymptotics~\cite{ott-mann:2000,biferale-etal:2005,ouellette-etal:2006,salazar-collins:2009,eyink:2011}.
Also, much work on relative dispersion has used it as a basis. For
instance, improvements of (\ref{eq:pdf_richardson}) were proposed
using modified versions of the eddy diffusivity $K(r)$, adding a time
dependence \cite{batchelor:1952,kraichnan:1966} or, more recently,
including finite Reynolds number effects
\cite{scatamacchia-etal:2012}. All of these improvements strongly
alter the functional form of the large-$r$ tail of the transition
probability density. Nevertheless, the physical mechanisms leading to
the de-correlation of velocity differences and to models based on eddy
diffusivity are still unclear and many questions remain open. For
instance, the speed of convergence to the Richardson--Obukhov law and
the form of subleading terms are still not known.

The first work dealing with the way pair separations converge to a
superdiffusive behavior is due to Batchelor \cite{batchelor:1950}.  He
argued that the explosive $t^3$ law is preceded by a ballistic phase
during which the tracers keep their initial velocity and separate
linearly in time, i.e.\ $R(t) - r_0 \simeq t\delta_{r_0} u \sim t
(\epsilon\,r_0)^{1/3}$, up to a time $\tau_{r_0} \sim \epsilon^{-1/3}
r^{2/3}_0$, which is equal to the eddy turnover time associated to the
initial separation. This first regime where velocity remains strongly
correlated can clearly not be described by eddy-diffusivity
approaches. Various stochastic models have been designed to catch the
two regimes. Most of them are based on the observation that the
acceleration difference between the two tracers is shortly correlated
and do not assume that velocity differences get uncorrelated. The pair
separation and the velocity difference can then be approximated as
coupled Markovian diffusive processes (see
\cite{kurbanmuradov:1997,sawford:2001} for reviews). The usual path
for designing such models requires imposing some constraints on the
drift and the diffusion terms.  Thomson \cite{thomson:1987} argued
that they should satisfy the well-mixed condition: when averaging over
uniformly separated pairs inside the whole inertial range, the
statistics of velocity differences between the two tracers has to
recover Eulerian two-point statistics.  Definitively devising an
admissible model requires an input from Eulerian statistics
\cite{borgas-yeung:2004}. It is known in turbulence that the
distribution of velocity differences is neither self-similar nor
Gaussian (see, e.g., \cite{frisch:1996}). Such models become thus so
complicated that they can hardly be used to improve the understanding
of the underlying phenomenology and, at the same time, they are not
easily amenable for an analytical treatment.

The goal of the present work is to give some new phenomenological
understanding of the problem of turbulent relative dispersion. For
this, we make use of large-scale direct numerical simulations and
essentially focus our study on the role of violent events leading to
strong fluctuations. To give grounds to possible improvements of
models for relative dispersion, we introduce some new observables,
measure their behavior, and interpret the results mostly in terms of
geometrical arguments.  The paper is organized as follows. We first
report in Sec.~\ref{sec:averaged_separations} numerical results on the
time evolution of averaged separations and of higher-order statistics;
this is an extension of some of our recent
work~\cite{bitane-etal:2012}. In Sec.~\ref{sec:veldiff} we focus on the
behavior of velocity differences and discuss the question of
intermittency and of a possible multiscaling for the time evolution of
their moments.  We also introduce a mixed moment of separation and
velocity difference that attains quickly a scaling regime (independent
on time and initial separations). In Sec.~\ref{sec:extreme}, we apply
previous results and make use of geometrical considerations to
investigate the very large and small fluctuations of the
inter-trajectory distance. Finally, Sec.~\ref{sec:conclusion}
encompasses concluding remarks and discuss open questions.

%%%%%%%%%%%%%%%%%%%%%%%%%%%%%%%%%%%%%%%%%%%%%%%%
\section{Convergence of separation statistics to a scaling regime}
\label{sec:averaged_separations}
%%%%%%%%%%%%%%%%%%%%%%%%%%%%%%%%%%%%%%%%%%%%%%%%

\subsection{Numerical results on the mean-square displacement}

We report in this section numerical results on the time evolution of
the averaged squared distance between tracers.  The aim is to provide
a first insight on the numerical simulations that we have used and on
the typical values of the timescales and lengthscales that are used
throughout this paper. One of the major difficulties encountered when
attacking numerically the problem of turbulent explosive separation is
that it requires a huge timescale separation. Indeed, to observe
Richardson--Obukhov $t^3$ superdiffusive regime, one needs to follow
particle pairs on a time $t$ much longer than the eddy turnover time
$\tau_{r_0} \sim \epsilon^{-1/3}r_0^{2/3}$ associated to their initial
separation $|\bR(0)|=r_0$ and much smaller than the time $\tau_L$
associated to the integral scale $L$. Also, it is in principle
required that the initial distance $r_0$ belongs to the inertial
range, so that its associated turnover time $\tau_{r_0}$ has to be
much larger than the Kolmogorov dissipative timescale $\tau_\eta =
(\nu/\epsilon)^{1/2}$. We should thus have $\tau_\eta\ll\tau_{r_0}\ll
t\ll \tau_L$. To give a concrete idea, if we require that these
timescales are separated by at least a decade, this implies that one
should have a Taylor-based Reynolds number $R_\lambda = \sqrt{15}
\,(\tau_L/\tau_\eta) \gtrsim 4000$. Such a high level of turbulence is
still far from the current state of direct numerical simulations and
also from present-day experimental setups where accurate particle
tracking techniques can be used (see
\cite{salazar-collins:2009,toschi-bodenschatz:2009} for recent
reviews). For this reason, there is an important need for
understanding the full process of convergence to the $t^3$ law in
order to predict and detect for instance possible subleading
corrections. This is the spirit in which we have tried to present this
work.

To investigate such issues, we have performed direct numerical
simulations of the incompressible Navier--Stokes equation in the
three-dimensional $2\pi$-periodic domain using a standard
pseudo-Fourier-spectral solver with a third-order Runge--Kutta time
marching. Such a method is well adapted to incompressible homogeneous
and isotropic turbulence at high Reynolds numbers with an extended
inertial range of scales. It has the advantages of combining a high
degree of accuracy with very good performances on massive parallel
supercomputers such as BlueGene systems and large Intel/AMD
clusters. We have used two sets of simulations whose parameters are
summarized in table~\ref{table} (more details on these simulations can
be found in \cite{grauer-etal:2010}). To maintain a statistical steady
state, the flow is forced by keeping constant the energy content of
the two first shells of wavenumbers in Fourier space.
\begin{table}[tbp]
  \tbl{Parameters of the numerical simulations}
  {\begin{tabular}{@{}ccccccccccc} \toprule
      $N$ & $R_\lambda$ &$\delta x$ & $\delta t$ & $\nu$ & $\epsilon$
      & $u_\mathrm{rms}$ &$\eta$&$\tau_\eta$&$L$&$\tau_L$\\\colrule
      $2048^3$ & $460$ & $3.7\!\cdot\!10^{-3}$ & $6\!\cdot\!10^{-4}$ & 
      $2.5\!\cdot\!10^{-5}$ &  $3.6\!\cdot\!10^{-3}$ & $0.19$ &
      $1.4\!\cdot\!10^{-3}$ & $0.083$ & $1.85$ & $9.9$ \\\colrule
      $4096^3$ & $730$ & $1.53\!\cdot\!10^{-3}$ & $1.2\!\cdot\!10^{-3}$ 
      & $1.0\!\cdot\!10^{-5}$ &  $3.8\!\cdot\!10^{-3}$ & $0.19$ & 
      $7.2\!\cdot\!10^{-4}$ & $0.05$ & $1.85$ & $9.6$ \\\botrule
    \end{tabular}}
  \tabnote{$N$ is the number of grid points, $R_\lambda$ the
    Taylor-based Reynolds number, $\delta x$ the grid
    spacing, $\delta t$ the time step, $\nu$ the kinematic viscosity, 
    $\epsilon$ the averaged energy dissipation rate, $u_\mathrm{rms}$
    the root-mean square velocity, $\eta \!=\! (\nu^3/\epsilon)^{1/4}$
    the Kolmogorov dissipative scale, $\tau_\eta \!=\!
    (\nu/\epsilon)^{1/2}$ the associated turnover time, $L \!=\!
    u_\mathrm{rms}^3/\epsilon$ the integral scale and $\tau_L \!=\!
    L/u_\mathrm{rms}$ the associated large-scale turnover time.}
\label{table}
\end{table}
\begin{figure}[h]
  \centering
  \includegraphics[width=\textwidth]{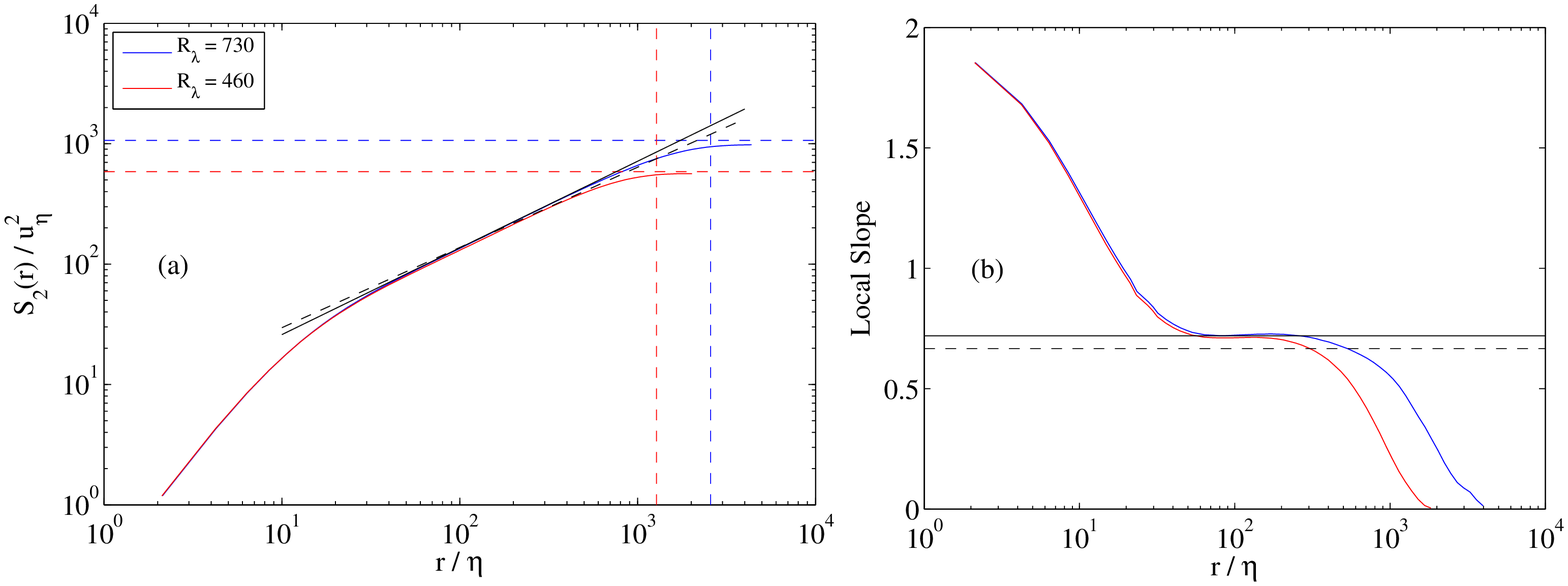}
  \caption{\label{fig:S2_2reynolds} (a) Second-order Eulerian structure
  function $S_2(r) = \langle |\bm u(\bm x+\bm r,t)-\bm u(\bm
  x,t)|^2\rangle$ for the two values of the Reynolds number
  investigated in this paper. The solid line represents Kolmogorov 1941
  scaling $S_2(r) \simeq (11/3)\,C_2 (\epsilon\,r)^{2/3}$ with $C_2=2.13$. The
  dashed line corresponds to She-L\'ev\^eque anomalous scaling with
  $\zeta_2 \approx  0.696$. The vertical and horizontal colored dashed
  lines indicate the integral scale and the large-scale velocity,
  respectively. (b) Logarithmic derivative $\mathrm{d}\log S_2
  /\mathrm{d}\log r$. The two horizontal lines encode the two scalings
  shown in Fig.~(a).}
\end{figure}
The Eulerian second-order structure function measured from these two
simulations are represented in Fig.~\ref{fig:S2_2reynolds}. As seen
there, the largest one develops a significant scaling range where
deviations from Kolmogorov 1941 scaling start to be visible.

For each value of the Reynolds number, the flow is seeded with $10^7$
tracer particles whose motion is integrated using tri-linear
interpolation. After a time sufficiently long to have converged to a
statistical steady state, we
start the analysis of the dispersion of tracer pairs. For this, we
label at a fixed initial time (that we fix here to be $t=0$) all
couples whose distance $|\bR(0)| = |\bm X_1(0) -\bm X_2(0)|$ is equal
to $r_0 \pm \eta\,$ for $\,r_0\le 16\,\eta$ and equal to $r_0\, (1\pm
2\%)$ for $r_0> 16\,\eta$ with $r_0 =
2,3,4,6,8,12,16,24,32,48,64,96,128,$ and $192\eta$. This bining was
chosen such that each family indexed by $r_0$ contains a few hundreds
of thousands of pairs. We then track forward in time all indexed pairs
and perform statistics conditioned on their initial separation
$r_0$. For the sake of simplicity we denote by $\langle\cdot\rangle$
the Lagrangian ensemble average conditioned on $r_0$.

\begin{figure}[h]
  \includegraphics[width=\textwidth]{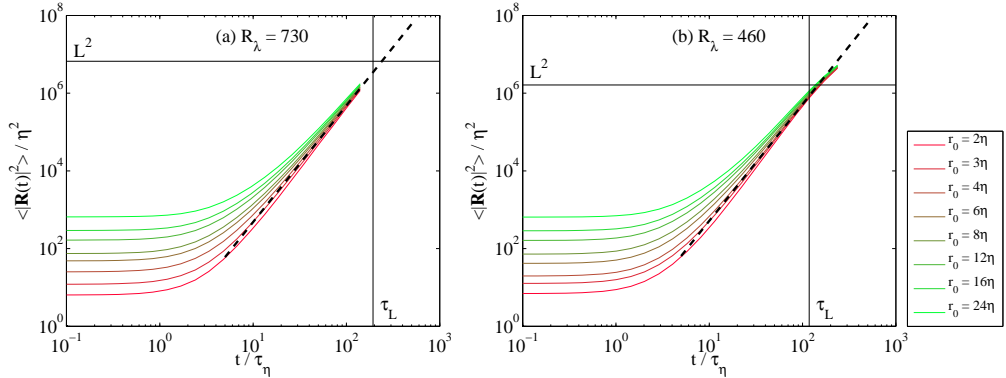}
  \caption{\label{fig:r2_fn_time_diffsep} Time-evolution of the
    mean-squared distance for $R_\lambda = 730$ (a) and $R_\lambda =
    460$ (b) for various initial separations $r_0$ as labeled. The
    horizontal and vertical solid lines represent the integral scale $L$ and
    its associated turnover time $\tau_L$, respectively. The dashed
    line corresponds to the explosive Richardson-Obukhov law
    (\ref{eq:explosive}) with $g=0.52$.}
\end{figure}
\begin{figure}[t]
  \includegraphics[width=0.5\textwidth]{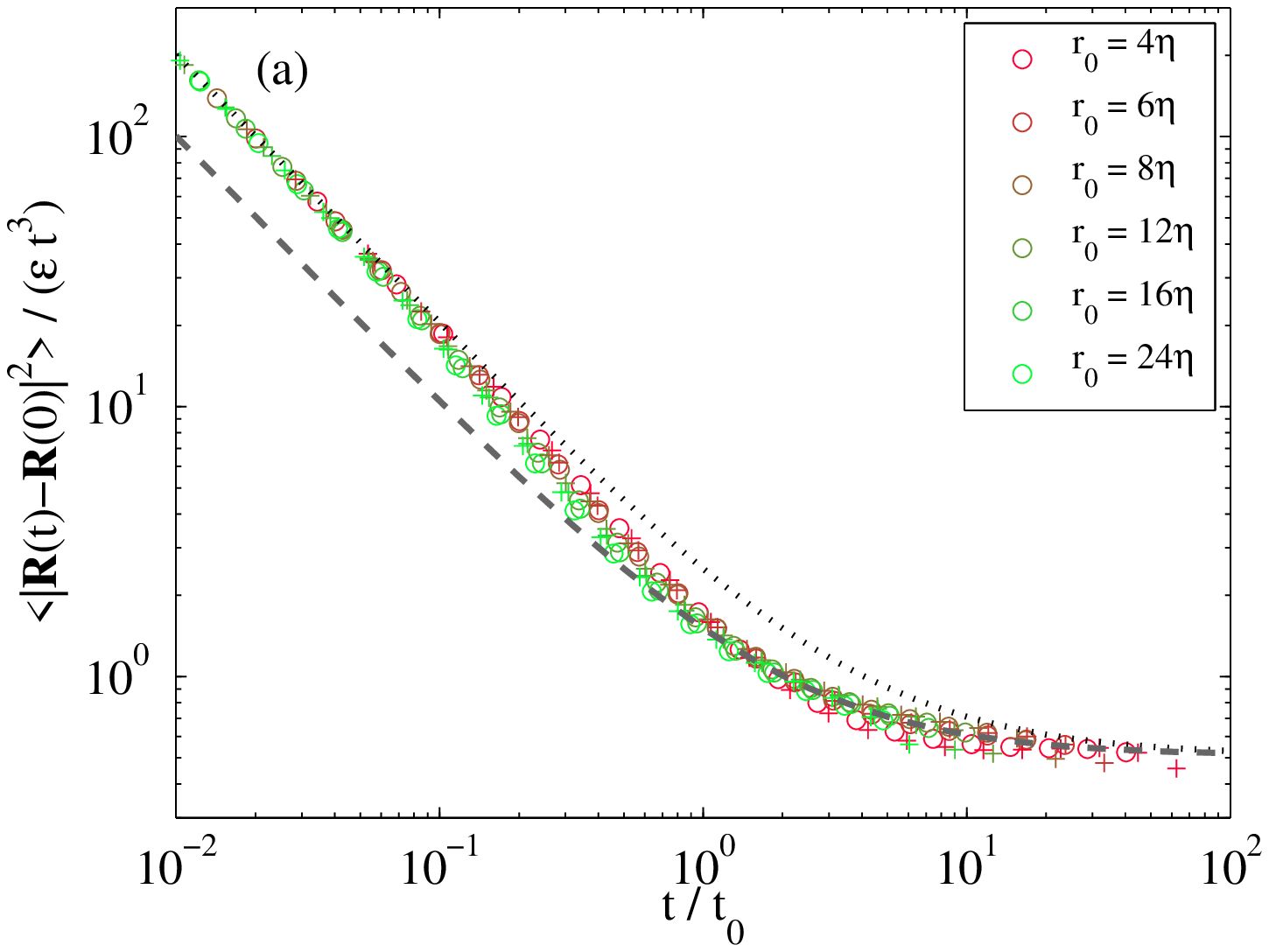}
  \hfill
  \includegraphics[width=0.5\textwidth]{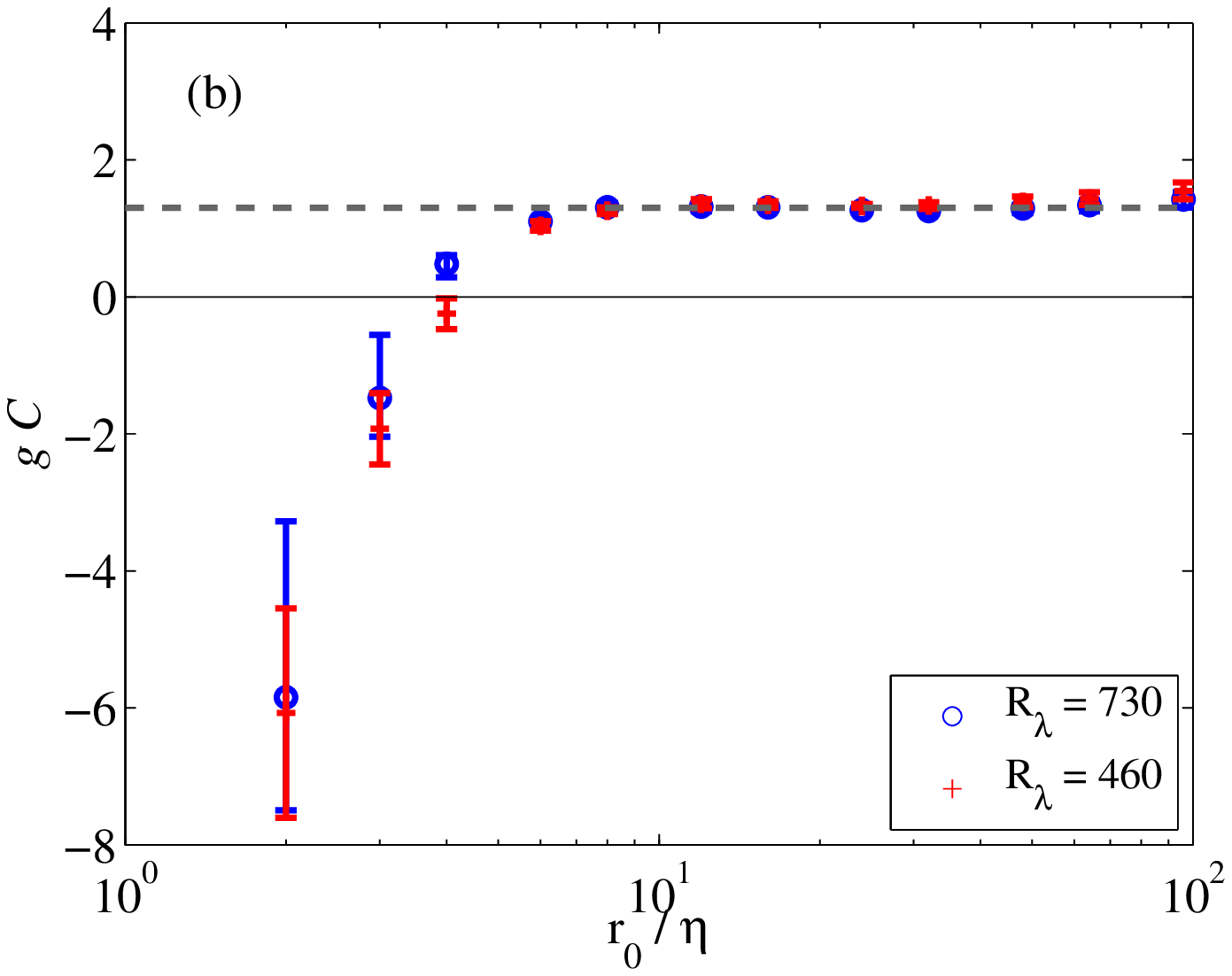}
  \caption{\label{fig:dr2_fn_time_compensate} (a) Compensated
    mean-squared displacement $\langle |\bR(t)-\bR(0)|^2\rangle /
    (\epsilon\,t^3)$ as a function of $t/t_0$ with $t_0 = S_2(r_0) /
    (2\epsilon)$ for various initial separations and $R_\lambda = 730$
    ($\circ$) and $R_\lambda = 460$ ($+$). The two curves show
    behaviors of the form $\langle |\bR(t)-\bR(0)|^2\rangle \simeq
    g\,\epsilon\,t^3 + A\,t^2$, with $A = S_2(r_0)$, given by
    Batchelor's ballistic regime (black dotted line), and $A =
    2.5/t_0^2$ (grey dashed line). (b) Measured value of the constant
    $g\,C$ in front of the subleading term as a function of the
    initial separation. It stabilizes to $C \approx 1.3/g \approx 2.5$
    for $r_0\gg\eta$; this value is represented as a dashed line. }
\end{figure}

Figure~\ref{fig:r2_fn_time_diffsep} shows for the two simulations the
time evolution of the mean squared distance $\langle |\bR(t)|^2
\rangle$ for various values of the initial separation $r_0$. Times and
space are there represented in dissipative-scale units. After a
transient (which roughly corresponds to Batchelor's ballistic regime),
the mean-squared distance approaches the explosive Richardson--Obukhov
regime
\begin{equation}
  \left\langle |\bR(t)|^2 \right\rangle \simeq g\,\epsilon\,t^3.
  \label{eq:explosive}
\end{equation}
We observe for both values of the Reynolds number a
Richardson--Obukhov constant $g\approx 0.52 \pm 0.05$. The low
accuracy with which this constant is determined comes from the fact
that, even at the higher resolution, the $t^3$ scaling is observed in
a rather limited time range. This is even clearer from
Fig.~\ref{fig:dr2_fn_time_compensate}~(a), which shows the compensated
mean squared increase of the distance $\langle
|\bR(t)-\bR(0)|^2\rangle / (\epsilon\,t^3)$. On this figure, the time
has been rescaled by $t_0 = S_2(r_0) / (2\epsilon)$, where $S_2$
designates the second-order Eulerian structure function with absolute
values. The choice of such a timescale was motivated in
\cite{bitane-etal:2012} as that of deviations from Batchelor's initial
ballistic regime. Surprisingly, the data shown in
Fig.~\ref{fig:dr2_fn_time_compensate}~(a) corresponding to various
initial separations $r_0$ far enough in the inertial range seem to
collapse.
This suggests that the timescale $t_0$ contains most of the dependence
of pair dispersion upon the initial separation $r_0$. Also data
indicate that the subdominant terms in (\ref{eq:explosive}) are
$\propto t^2$, leading to postulate $\left\langle |\bR(t)|^2
\right\rangle \simeq g\,\epsilon\,t^3 (1+C\,t_0/t)$, with a constant
$C$ independent of $r_0$ when $r_0\gg\eta$. The product of the
constants $g$ and $C$ has been estimated numerically and results are
shown in Fig.~\ref{fig:dr2_fn_time_compensate}~(b). One can clearly
see that $C \approx 1.3/g \approx 2.5$ when $r_0\gtrsim 8\eta$.

One can also see from the figure that $C<0$ for $r_0\lesssim
4\eta$. Thus, for dissipative-range initial separations, the
asymptotic $t^3$ behavior is attained from below. This can lead for
such values of $r_0$ to an intermediate time range where the mean
squared distance grows even faster than the explosive $t^3$ law, as
for instance observed in \cite{biferale-etal:2005}.  Another remark
that can be drawn from the data is that, independently of the Reynolds
number, the constant $C$ is equal to zero for $r_0\approx 4\eta$. The
first subleading terms are then $\propto t$, so that the convergence
to the $t^3$ law is much faster for such an initial separation than
for others. This observation could be useful for experimentalists to
optimize their setup. However, such small values of $r_0$ are clearly
not representative of the inertial-range behavior.

%%%%%%%%%%%%%%%%%%%%%%%%%%%%%%%%%%%%%%%%%%%%%%%%
\subsection{Higher-order statistics}
\label{subsec:higher-order}

\begin{figure}[h]
  \includegraphics[width=\textwidth]{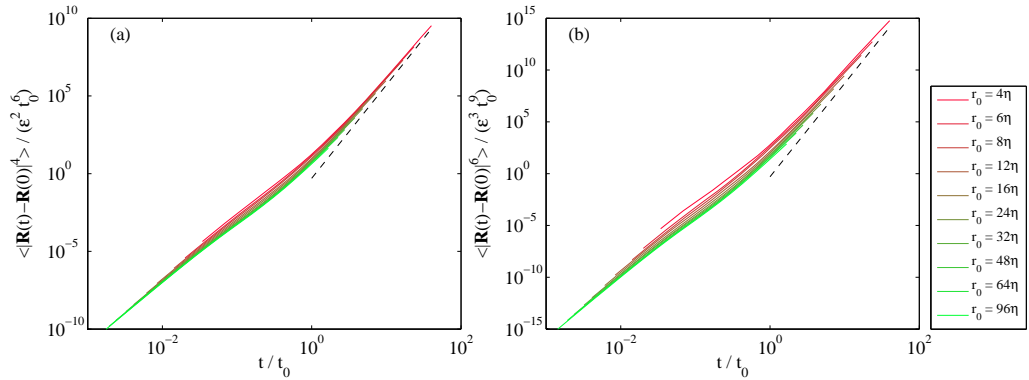}
  \caption{\label{fig:r4_fn_time_diffsep} (a) Fourth-order moment
    $\langle|\bR(t) -\bR(0)|^4\rangle$ and (b) sixth-order moment
    $\langle|\bR(t) -\bR(0)|^6\rangle$ as function of $t/t_0$ for
    $R_\lambda = 730$. Both curves are normalized such that their
    expected long-time behavior is $\propto(t/t_0)^6$ and
    $\propto(t/t_0)^9$, respectively. The black dashed lines represent
    such behaviors. }
\end{figure}
We now turn to investigating the large-time behavior of higher-order
moments of the separation. Figure \ref{fig:r4_fn_time_diffsep} shows
the time evolution of $\langle|\bR(t)-\bR(0)|^4\rangle$ (a) and of
$\langle|\bR(t)-\bR(0)|^6\rangle$ (b).  At times smaller than $t_0$
the separation grows ballistically, so that $\langle|\bR(t)-\bR(0)|^p
\rangle \simeq t^p\, \langle|\bV(0)|^p\rangle$ where $\bV(t) \,=\, \bm
u(\bm X_1,t)\,-\,\bm u(\bm X_1,t)$ denotes the velocity difference
between the two tracers. The fact that we have chosen to rescale time
by $t_0$ (which depends on second-order statistics of the initial
velocity difference) implies that the moments do not collapse in this
regime because of Eulerian multiscaling. However the collapse occurs
for $t\gg t_0$ where these two moments grow like $t^6$ and $t^9$,
respectively, with possible minute deviations. The measured power-laws
give evidence that, at sufficiently long times, inter-tracer distances
follow a scale-invariant law. Also the observed collapses indicate
that $t_0$ could be again the time of convergence to such a behavior.

\begin{figure}[h]
  \includegraphics[width=\textwidth]{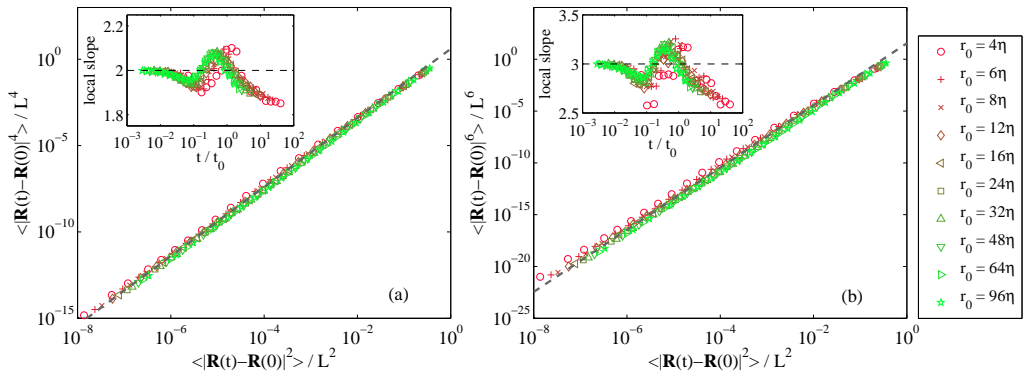}
  \caption{\label{fig:R_no_intermittency} Fourth (a) and sixth (b)
    order moments of $|\bR(t) - \bR(0)|$ as a function of its
    second-order moment for $R_\lambda = 730$. The two gray dashed
    lines show a scale-invariant behavior, i.e.\
    $\langle|\bR(t)\!-\!\bR(0)|^4\rangle \propto \langle|\bR(t)
    \!-\!\bR(0)|^2\rangle^2$ and $\langle|\bR(t) \!-\!\bR(0)|^6\rangle
    \propto \langle|\bR(t) \!-\!\bR(0)|^2\rangle^3$, respectively. The
    two insets show the associated local slopes, that is the logarithmic
    derivatives $\mathrm{d}\log \langle|\bR(t)\!-\!\bR(0)|^p\rangle /
    \mathrm{d}\log \langle|\bR(t)\!-\!\bR(0)|^2\rangle$, together with
    the normal scalings represented as dashed lines.}
\end{figure}
The presence of a scale-invariant regime is also clear when making use
of ideas borrowed from extended self-similarity and representing these
two moments as a function of $\langle|\bR(t)-\bR(0)|^2\rangle$ (see
Fig.~\ref{fig:R_no_intermittency}). This time, for a fixed $r_0$, the
smallest separations correspond to the ballistic regime. There, we
trivially have $\langle|\bR(t) \!-\!\bR(0)|^p\rangle
/\langle|\bR(t)\!-\!\bR(0)|^2 \rangle^{p/2} \simeq
\langle|\bV(0)|^p\rangle/\langle|\bV(0)|^2\rangle^{p/2} $, which has a
weak dependence on $r_0$, because of an intermittent distribution of
Eulerian velocity increments, but does not depend on time. This normal
scaling can be observed for $t\ll t_0$ in the insets of
Fig.~\ref{fig:R_no_intermittency}, which represent the logarithmic
derivatives of the high-order moments with respect to the second
order. At times of the order of $t_0$, noticeable deviations to normal
scaling can be observed. Finally, at much larger scales, data
corresponding to different values of the initial separation $r_0$
collapse but the curves start to bend down.  One observes in the
insets that the associated local slopes approach values clearly
smaller than those corresponding to normal scaling.  This gives
evidence of a rather weak intermittency in the distribution of tracer
separations. Note that the presented measurements were performed for
$R_\lambda = 730$ but the same behavior has been observed for
$R_\lambda = 460$.

To our knowledge, the most convincing observation of an intermittent
behavior in pair dispersion has been based on an exit-time
analysis~\cite{boffetta-sokolov:2002b}. However, the relation of such
fixed-scale statistics to the usual fixed-time measurements we report
here requires to consider pair separation velocities. As we will see
in next Section, the velocity difference between two tracers displays
statistics that are much more intermittent than those for pair
separation. This implies that there is no contradiction between an
almost normal scaling for distances as a function of time and an
anomalous behavior of exit times as a function of distance.

\begin{figure}[h]
  \includegraphics[width=0.5\textwidth]{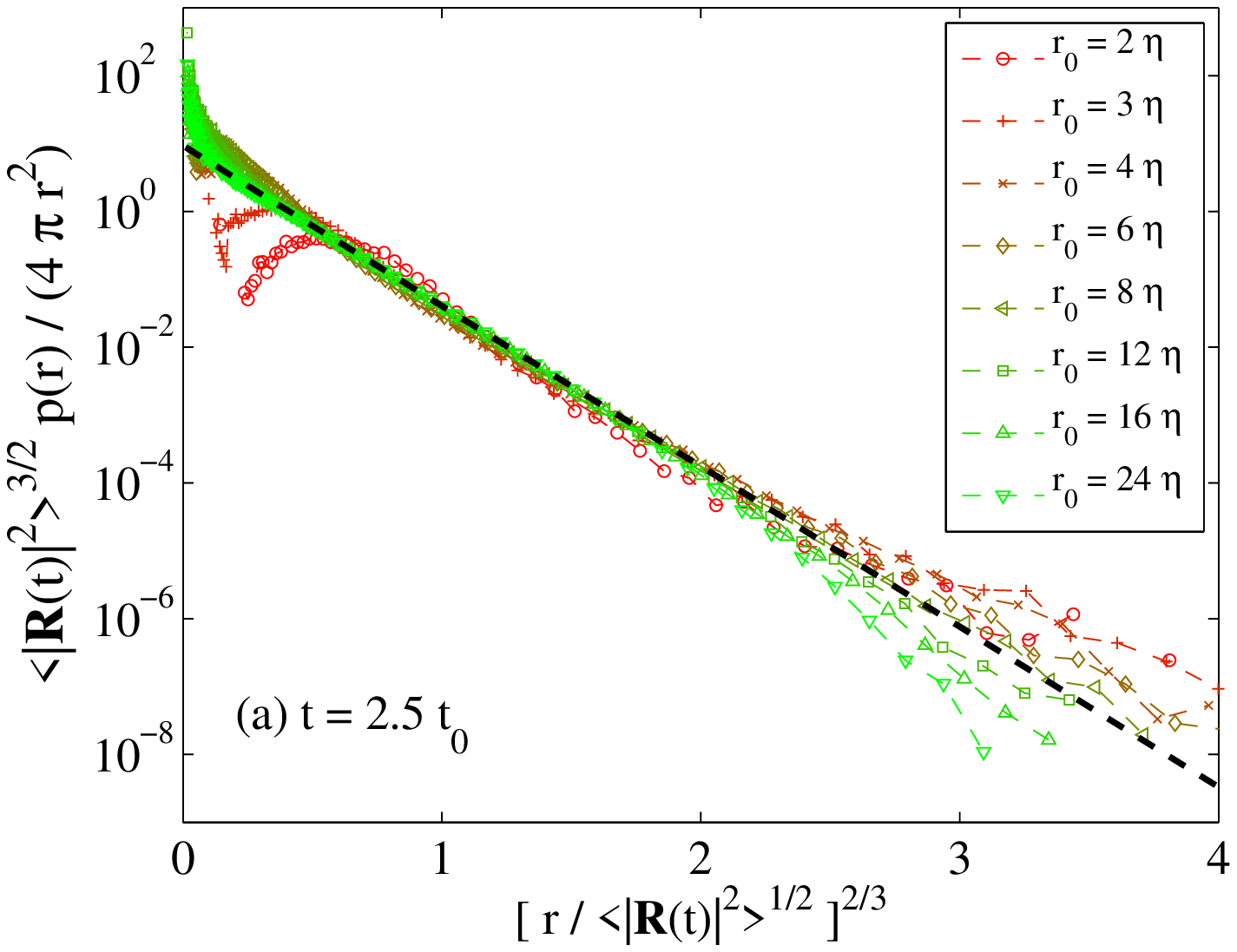}
  \hfill
  \includegraphics[width=0.5\textwidth]{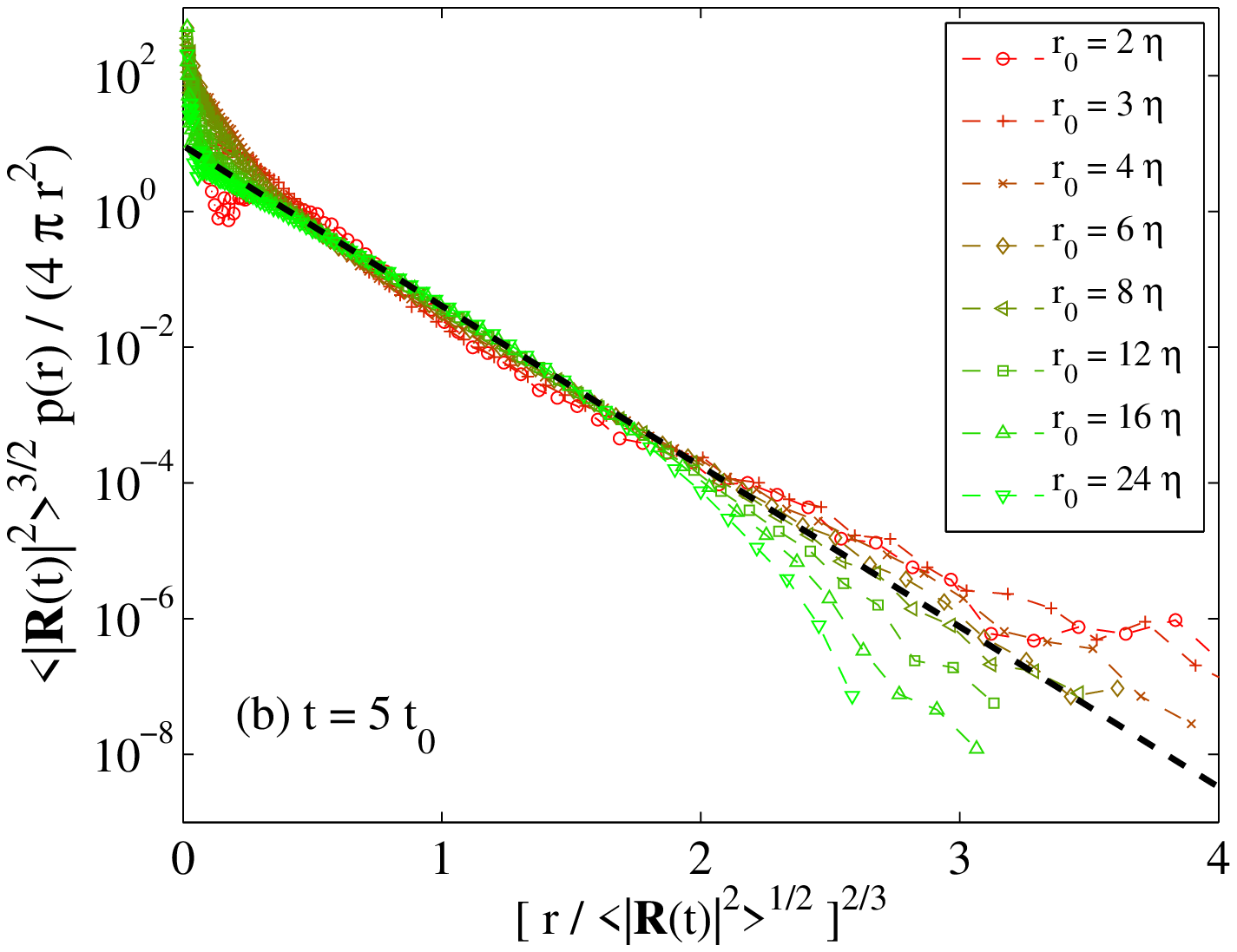}
  \caption{\label{fig:histo_R_4096} Probability density function of
    the distance $r$ at time $t = 2.5\,t_0$ (a) and $t = 5\,t_0$ (b)
    and for various values of the initial separation. We have here
    normalized it by $4\pi r^2$ and represented on a $\log y$ axis as a
    function of $r/\langle |\bR(t)|^2\rangle^{1/2}$. With such a
    choice, Richardson's diffusive density distribution
    (\ref{eq:pdf_richardson}) appears as a straight line (represented
    here as a black dashed line).}
\end{figure}
To investigate further this weak intermittency in the separation
distribution, we have represented in Fig.~\ref{fig:histo_R_4096} the
probability density function (PDF) of the distance $|\bR(t)|$ for
various initial separation and at times where we expect to have almost
converged to the explosive regime, namely at $t=2.5\,t_0$ (a) and
$t=5\,t_0$ (b). Such measurements are compared to Richardson's
diffusive law (\ref{eq:pdf_richardson}). Data suggest that a large
part of the PDF core (for $0.4\lesssim |\bR(t)|/\langle
|\bR(t)|^2\rangle^{1/2}\lesssim 4$ at time $t=5\,t_0$) is very well
described by Richardson's approach.  However, deviations are observed
in the far tails, at both small and large values of the
separation. Such an observation is consistent with previous numerical
observations
\cite{boffetta-sokolov:2002b,biferale-etal:2005,scatamacchia-etal:2012}. Apparently,
these deviations affect only weakly the moments we have considered
above. We will come back to investigating and characterizing them in
Sec.~\ref{sec:extreme}.

%%%%%%%%%%%%%%%%%%%%%%%%%%%%%%%%%%%%%%%%%%%%%%%%
\section{Statistics of velocity differences}
\label{sec:veldiff}
%%%%%%%%%%%%%%%%%%%%%%%%%%%%%%%%%%%%%%%%%%%%%%%%
\subsection{A diffusive behavior?}

In this section, we are interested in the behavior of the velocity
difference $\bV(t) = \bm u(\bm X_1(t),t) - \bm u(\bm X_2(t),t)$
between two tracers as a function of time. Initially, the statistics
of $\bV(0)$ are exactly given by the Eulerian statistics of velocity
increments at a separation $r_0$. At large times, an explosive $t^3$
behavior for distances implies that $\langle |\bV(t)|^2 \rangle
\propto \epsilon\, t$. A naive picture would consist in interpolating
between these two behaviors by assuming that
\begin{equation}
  \langle |\bV(t)|^2 \rangle \simeq S_2(r_0) + h\,\epsilon\, t,
  \label{eq:behav_vel}
\end{equation}
where $h$ is a positive constant that cannot be straightforwardly
expressed as a function of $g$. As argued in \cite{bitane-etal:2012}
one of the two terms, which correspond to the ballistic or to the
Richardson--Obukhov regime, is dominant when $t$ is much smaller or
much longer than $t_0 = S_2(r_0)/(2\,\epsilon)$. 
\begin{figure}[b]
  \includegraphics[width=0.5\textwidth]{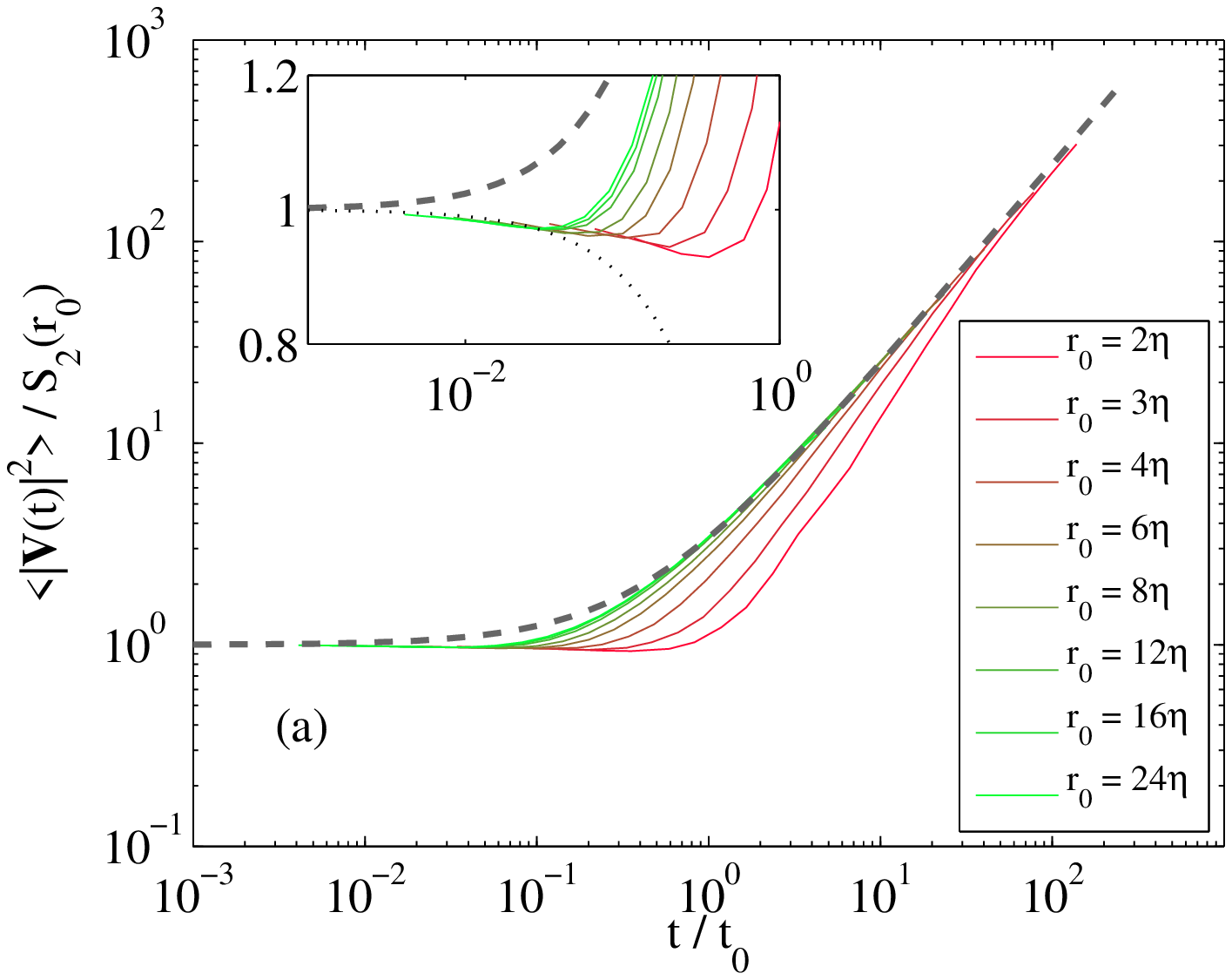}
  \hfill
  \includegraphics[width=0.5\textwidth]{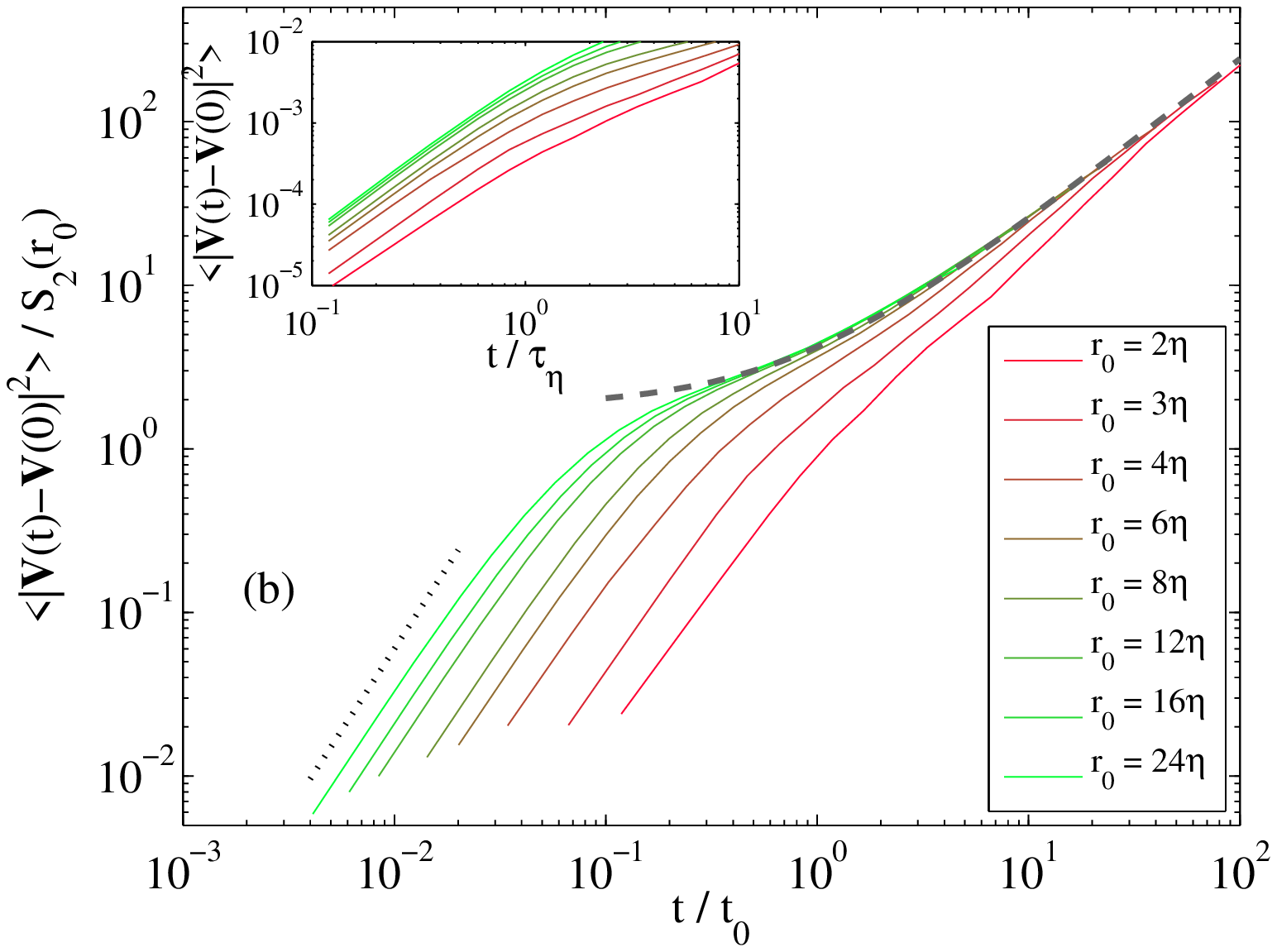}
  \caption{\label{fig:dv2_fn_time} Time behavior of the mean-squared
    velocity difference for $R_\lambda = 730$. (a) $\langle
    |\bV(t)|^2\rangle$, normalized by its initial value, as a function
    of $t/t_0$; the gray dashed line is a behavior of the form
    $\langle |\bV(t)|^2 \rangle \simeq S_2(r_0) + h\,\epsilon\, t$
    with $h = 2.4$. Inset: same but zoomed at short times; the black
    dotted line is the initial decrease associated to kinetic energy
    dissipation: $\langle |\bV(t)|^2 \rangle \simeq S_2(r_0) -
    2\,\epsilon\,t$. (b) Time behavior of the velocity change $\langle
    |\bV(t) - \bV(0)|^2 \rangle$; the dashed curve is $\langle
    |\bV(t)- \bV(0)|^2 \rangle \simeq S_2(r_0) + h\,\epsilon\, t$; the
    dotted line has slope $2$ and corresponds to the initial abrupt
    variation of the velocity difference. Inset: same but rescaling time
    by $\tau_\eta$.}
\end{figure}
As seen from Fig.~\ref{fig:dv2_fn_time}~(a) such a form with $h=2.4$
(represented as a dashed line) seems to be a good first-order
approximation. However, this picture is not completely
satisfactory. When zooming closer to the initial times, one observes
that the averaged pair kinetic energy $\langle |\bV(t)|^2 \rangle$
first starts by decreasing for a time of the order of a few tens of
$t_0$. We indeed know that initially
\begin{equation}
  \bV(t) = \bV(0) + t {\bm A}(0) + \mathcal{O}(t^2),
  \label{eq:ballistic}
\end{equation}
where ${\bm A}(0)$ denotes the initial difference of acceleration
between the two tracers. This approximation leads to $\langle
|\bV(t)|^2 \rangle \simeq S_2(r_0) - 2\,\epsilon\,t$ (see, e.g.,
\cite{bitane-etal:2012}).  According to our numerical data, this
regime seems to reproduce well what is happening for times up to
$\approx 0.01\,t_0$\,---\,see inset of
Fig.~\ref{fig:dv2_fn_time}~(a). The fact that this prediction ceases
to describe the data after such a short timescale (that is
significantly smaller than $t_0$ instead of being of its order)
indicates that the subleading terms in (\ref{eq:ballistic}) then
become important. As a consequence, depending on the quantity we are
interested in, Batchelor's ballistic regime might end very quickly.

After this subtle initial decrease, the averaged squared velocity
difference amplitude tends asymptotically to a behavior $\propto
t$. However this process occurs in a noticeably different manner to
that of convergence of squared separation to Richardson $t^3$ law:
irrespective of the initial separation $r_0$, the quantity $\langle
|\bV(t)|^2\rangle$ always approaches from below the asymptotic
behavior. In other terms, we observe that for a fixed value of
$t/t_0$, the pair kinetic energy $\langle |\bV(t)|^2\rangle$ is an
increasing function of $r_0/\eta$ and seems to converge to
(\ref{eq:behav_vel}) only in the limit $r_0/\eta\to\infty$.

At large times, the mean-squared velocity difference grows
linearly. This behavior, together with the fact that acceleration
differences are correlated over timescales of the order of the
Kolmogorov dissipative time $\tau_\eta$, suggest that velocity
differences have a diffusive behavior for $t\gg\tau_\eta$ (see
\cite{bitane-etal:2012}). If this was exact, it would lead to the
behavior (\ref{eq:behav_vel}) as the temporal increments of $\bV$
would be independent of its initial value. Also, a purely diffusive
behavior of $\bV$ would result in the fact that its mean-squared
temporal increment $\langle |\bV(t) \!-\! \bV(0) |^2\rangle$ is
$\propto t$ for all times $t\gg\tau_\eta$. As seen in
Fig.~\ref{fig:dv2_fn_time}~(b), the diffusive regime is actually only
asymptotic. Also, data show that $\bV(t)$ initially changes on times
of the order of $\tau_\eta$ by a factor of the order of its typical
initial value $[S_2(r_0)]^{1/2}$. This abrupt evolution can be
interpreted phenomenologically. With some finite probability, one of
the two tracers is within a vortex filament at time $t_0$. The typical
energy content of this filament will contribute to the value of
$S_2(r_0)$. However, after a time $t$ of the order of $\tau_\eta$, the
trajectory of this tracer will have turned around this filament, so
that its velocity will have completely changed orientation (without
changing much its amplitude). This will result in $|\bV(t) \!-\!
\bV(0) | \sim |\bV(t)|$, explaining the observed behavior. As a result
of this sudden kinematic variations of velocity differences, $\langle
|\bV(t)\!  -\!\bV(0) |^2\rangle$ behaves in a very similar manner to
$\langle |\bV(t)|^2\rangle$ for times $t\gg\tau_\eta$ (compare the
dashed lines in Fig.~\ref{fig:dv2_fn_time}~(a) and (b)). Let us also
notice that the convergence to this behavior is again from below,
irrespective of the initial separation $r_0$.

There is hence an abrupt change (occurring on timescales of the
order of $\tau_\eta$) that prevents from determining an effective
initial velocity difference and thus from observing a clear diffusive
behavior of $\bV(t)$. However, data suggest that the timescale of
convergence to this behavior is, as for separations, of the order of
$t_0$.  To understand further this question, we next turn to
investigating the behavior of the longitudinal velocity difference
between the tracers.

%%%%%%%%%%%%%%%%%%%%%%%%%%%%%%%%%%%%%%%%%%%%%%%%
\subsection{Geometry of longitudinal velocities}
\label{subsec:geom_long}

\begin{figure}[t]
  \includegraphics[width=\textwidth]{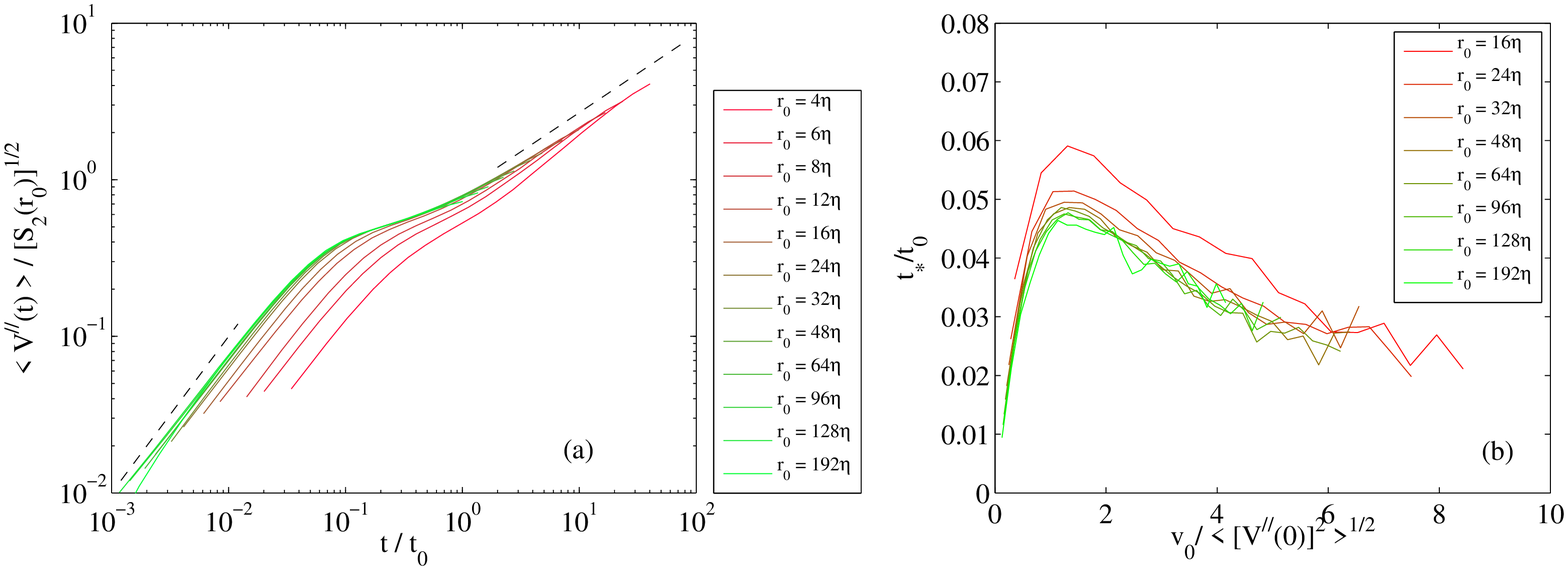}
  \caption{\label{fig:vpar_fn_time} (a) Time evolution of the averaged
    longitudinal velocity difference $\langle V^\parallel(t) \rangle$
    for various initial separations, as labeled, and for
    $R_\lambda=730$. (b) Averaged time $t_\ast = -r_0 v_0 / \langle
    |\bV(0)|^2\,|\,V^\parallel(0)=v_0\rangle$ when trajectories with a
    given initial longitudinal velocity difference $v_0$ reach a
    minimal distance.}
\end{figure}

We are here interested in the evolution of the longitudinal component
$V^\parallel(t) = \bR(t)\cdot\bV(t)/|\bR(t)|$ of the velocity
difference as a function of time. This quantity is important to
characterize pair separation as $\mathrm{d}|\bR|/\mathrm{d}t =
V^\parallel$. Initially, the averaged longitudinal velocity vanishes,
i.e.\ $\langle V^\parallel(0) \rangle =0$; this is due to the
statistical stationarity of the fluid flow. For times $t\ll t_0$ in
the Batchelor's ballistic regime, the pairs are keeping their initial
velocity difference and one can easily check that
\begin{equation}
  \langle V^\parallel(t) \rangle \simeq t\,\langle
  |\bV^\perp(t)|^2 \rangle/r_0,
\end{equation}
where $\bV^\perp$ denotes the components of $\bV$ that are transverse
to $\bR$. It is thus clear that the average velocity at which tracer
trajectories separate immediately becomes positive.
Figure~\ref{fig:vpar_fn_time}~(a), which represents the time evolution
of $\langle V^\parallel(t) \rangle$, shows without doubt this initial
linear growth.
This increase of the longitudinal velocity difference has an
interesting geometrical interpretation.  If all pairs were to maintain
indefinitely their initial velocity differences $\bV(0)$, it is clear
that they would eventually all separate at large times. Indeed, they
will reach a minimal distance at a finite time $t_\ast =
-[\bR(0)\!\cdot\!\bV(0)] / |\bV(0)|^2 = -r_0 V^\parallel(0) /
|\bV(0)|^2$, which is positive for particles that are initially
approaching. After the time $t_\ast$, the distance between particles
increases and $V^\parallel$ becomes positive. This leads to an
increase of $\langle V^\parallel\rangle$ that comes from kinematic
considerations and is obviously not due to any dynamics imposed by the
turbulent flow. To estimate the typical value of the minimal distance
at time $t_\ast$, we have performed statistics on pair separation
conditioned not only on the initial distance $r_0$, but also on the
initial longitudinal velocity difference by binning pairs with
$V^\parallel(0) = v_0 \pm \delta v_0$. From these statistics we have
defined the averaged time $t_\ast = -r_0 v_0 / \langle
|\bV(0)|^2\,|\,v_0\rangle$ at which trajectories with a given initial
longitudinal velocity difference $v_0$ are at a minimal distance. The
data are shown in Fig.~\ref{fig:vpar_fn_time} (b) for various values of
$r_0$ and as a function of $v_0$. We observe that for $r_0\gg\eta$
this time approximatively takes the form $t_\ast \simeq
t_0\,f(v_0/\langle |\bV(0)|^2\rangle^{1/2})$, where the function
$f(x)$ attains its maximum (roughly equal to 0.05) at $x\approx
1$. This confirms the observation made in Fig.~\ref{fig:vpar_fn_time}
(a) that the initial growth of $\langle V^\parallel\rangle$ occurs on
a time length of the order of a few hundredths of $t_0$.

\begin{figure}[h]
  \includegraphics[width=\textwidth]{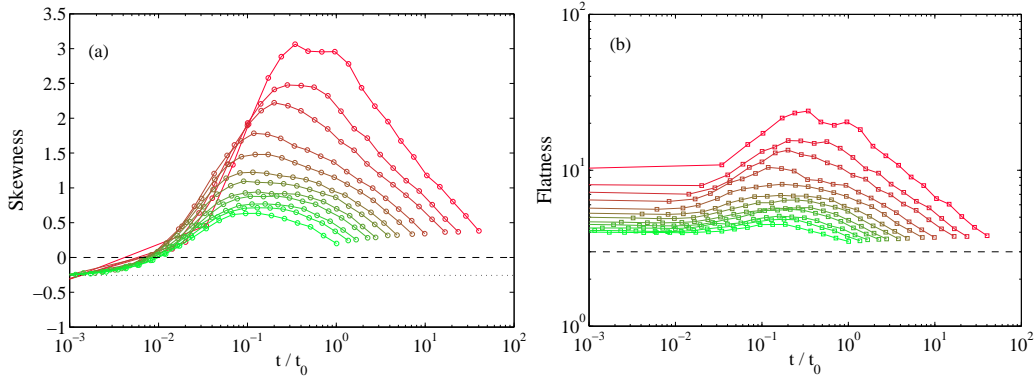}
  \caption{\label{fig:skew_flat} Time evolution of (a) the
    skewness $\mathcal{S}(t)$ and (b) the flatness $\mathcal{F}(t)$
    defined in (\ref{def:skew_flat}) for the same initial separations
    $r_0$ as in Fig.~\ref{fig:vpar_fn_time}~(a). The dashed lines are
    the values corresponding to a Gaussian distribution:
    $\mathcal{S}=0$ and $\mathcal{F}=3$. The dotted line in (a) is the
    initial value of the skewness obtained when assuming Kolmogorov
    1941 scaling, namely $\mathcal{S}= (4/5)/C_2^{3/2}$.}
\end{figure}

In addition to the change in the mean longitudinal velocity difference
discussed above, numerical measurements show that the full
distribution of $V^\parallel$ looses its symmetry and develops fatter
tails when time increases. Figures~\ref{fig:skew_flat} (a) and (b)
represent the skewness $\mathcal{S}$ and flatness $\mathcal{F}$ of
$V^\parallel$ as a function of time and for the same initial
separations as in Fig.~\ref{fig:vpar_fn_time} (a). These observables
are frequently used in turbulence to quantify the shape of the
velocity increment distribution. For two-particle Lagrangian
statistics, they are defined as
\begin{equation}
  \mathcal{S}(t) = \frac{\left\langle [V^\parallel(t)-\langle
      V^\parallel(t) \rangle]^3 \right\rangle}{\left\langle [V^\parallel(t)-\langle
      V^\parallel(t) \rangle]^2
    \right\rangle^{3/2}}\quad\mbox{and}\quad
  \mathcal{F}(t) = \frac{\left\langle [V^\parallel(t)-\langle
      V^\parallel(t) \rangle]^4 \right\rangle}{\left\langle [V^\parallel(t)-\langle
      V^\parallel(t) \rangle]^2 \right\rangle^{2}}.
  \label{def:skew_flat}
\end{equation}
As already observed for instance in \cite{yeung-borgas:2004}, these
quantities strongly vary as a function of time and maintain a marked
dependence upon the initial separation $r_0$ for rather long
times. Figure~\ref{fig:skew_flat}~(a) shows that the skewness of
$V^\parallel(t)$ starts from negative values (to be in agreement with
the Eulerian 4/5 law) and becomes positive at times larger than
$\approx 0.01\,t_0$. This initial change of sign can also be
interpreted geometrically in terms of the time $t_\ast$ when initially
ballistically approaching pairs begin to move away. However, after
this, the curves separate and each of them attains a maximum at times
of the order of $t_0$ or slightly smaller. This maximal value of the
skewness strongly depends on the initial separation: less is $r_0$,
higher it is. After this maximum, the skewness decreases without
attaining an asymptotic regime that would be independent of
$r_0$. This could be either due to the fact that there is a persistent
memory of $r_0$ in such quantities or to a contamination by finite
Reynolds number (and finite size) effects. The same kind of behavior
is observed for the flatness $\mathcal{F}$ of the distribution of
$V^\parallel$ as seen in Fig.~\ref{fig:skew_flat}~(b). However, the
increase of its maximal value when decreasing $r_0$ is even more
pronounced. Another noticeable difference is that the initial value of
$\mathcal{F}$ itself depends on $r_0$ and relates to the scale
dependence of the Eulerian flatness.

\begin{figure}[h]
  \includegraphics[width=\textwidth]{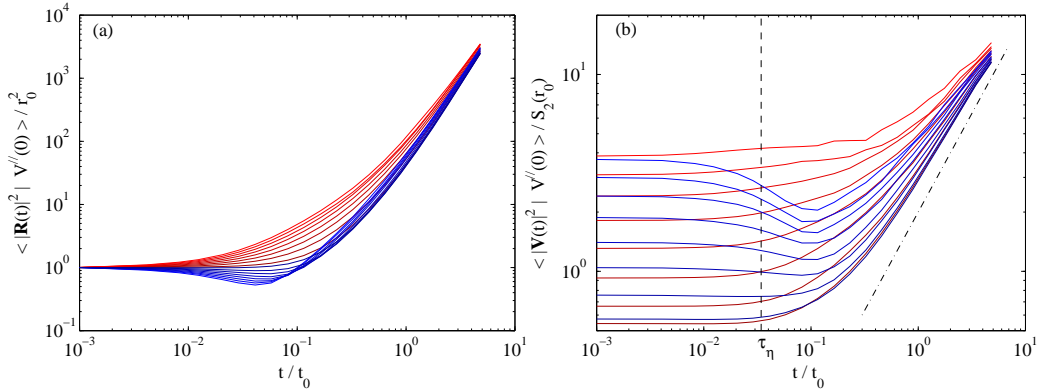}
  \caption{\label{fig:R2_V2_condvelpar} Time evolution for $r_0 =
    24\eta$ of the averaged distance (a) and of the squared velocity
    difference (b) conditioned on $V^\parallel(0) = v_0$ with $|v_0|
    /\langle [V^\parallel(0)]^2\rangle^{1/2} =
    0.1,0.4,0.8,1.2,...,2.8$. The red curves correspond to positive
    values of $v_0$ and the blue ones to negative values. In (b) the
    dashed-dotted line is $\propto t$}
\end{figure}
The strong dependence on $r_0$ of the skewness and of the flatness can
be interpreted phenomenologically in terms of the intermittent nature
of velocity increments. A large positive value of $\mathcal{S}$
corresponds to a large probability of having pairs separating faster
than the average. Such events will also be responsible for a large
value of the flatness $\mathcal{F}$. It is clear that those particles
that separate the faster are typically those which are the most
separated. Also, in a turbulent flow, the typical value of the
velocity increases with scale, so that the particles which get quickly
separated are likely to continue separating faster than the
average. This is evidenced in Fig.~\ref{fig:R2_V2_condvelpar}, which
represents the mean squared distance and the averaged squared
amplitude of the velocity difference conditioned on the initial value
of the longitudinal velocity difference for $r_0=24\eta$. One observes
that there is up to the latest time (of the order of $5\,t_0$), a
noticeable memory on the initial value of $V^\parallel$.  Besides this
consideration, one also remarks in Fig.~\ref{fig:R2_V2_condvelpar}~(b)
that the pairs having an initially large negative longitudinal
velocity difference (the blue curves) dissipate much more kinetic
energy than the others. However, this does not prevent them from
separating at large times faster than the pairs having initially a
smaller velocity difference. We now turn back to the explanation of
the long-term dependence on $r_0$ of the skewness and flatness of
$V^\parallel$. Recall that, in a turbulent flow, violent velocity
differences are more probable at small scales than at larger
scales. This implies that pairs with a small initial separation are
more likely to experience a large (positive or negative) initial
velocity difference. This will make them separate faster than the
average and thus experience even larger values of the velocity. The
rapid and strong increase of fluctuations in their velocity
differences is thus due to a kind of snowball effect. We will come
back later to quantifying with more accuracy this phenomenon.

%%%%%%%%%%%%%%%%%%%%%%%%%%%%%%%%%%%%%%%%%%%%%%%%
\subsection{Intermittency}
\begin{figure}[h]
  \includegraphics[width=0.5\textwidth]{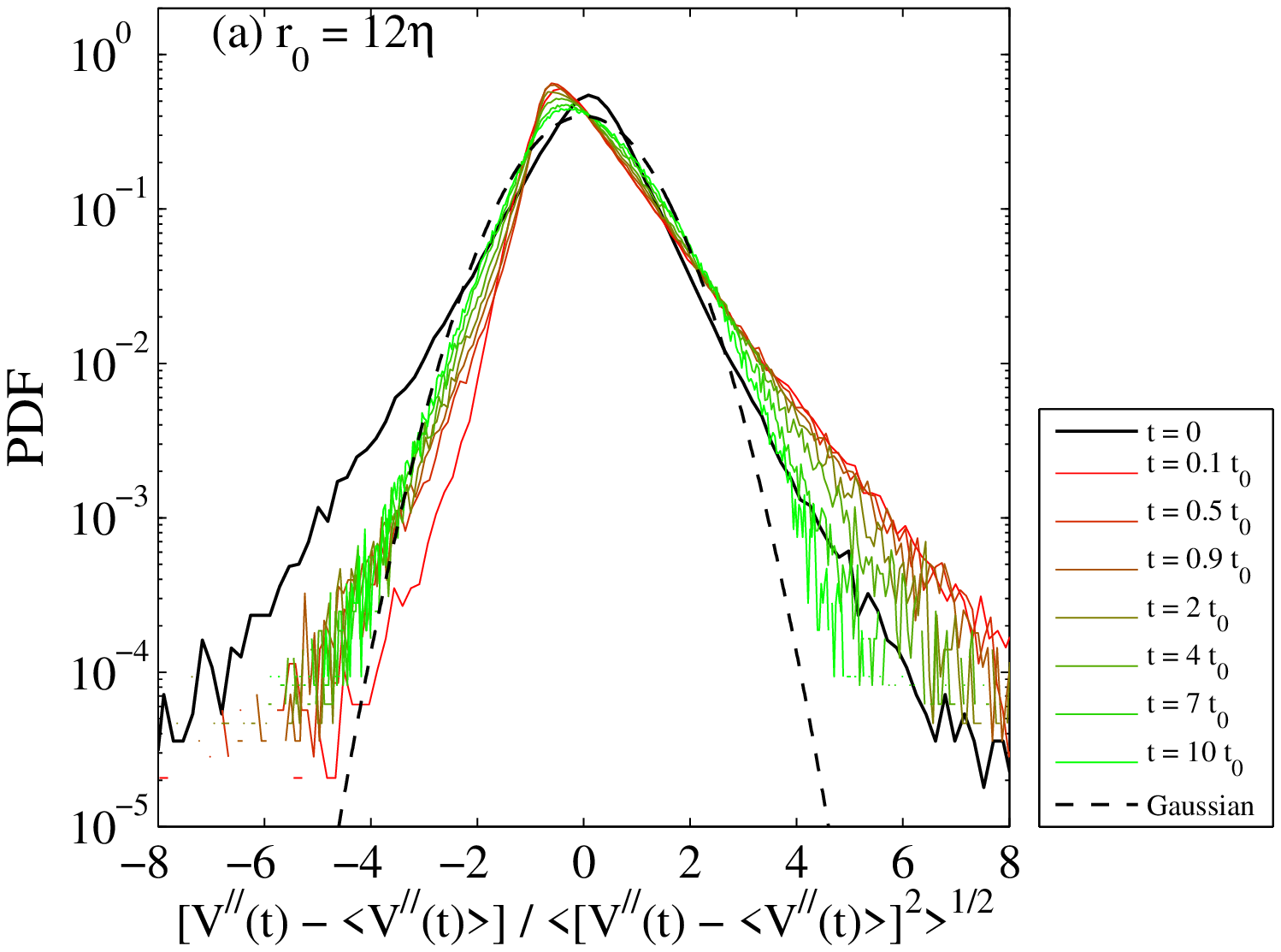}
  \hfill
  \includegraphics[width=0.5\textwidth]{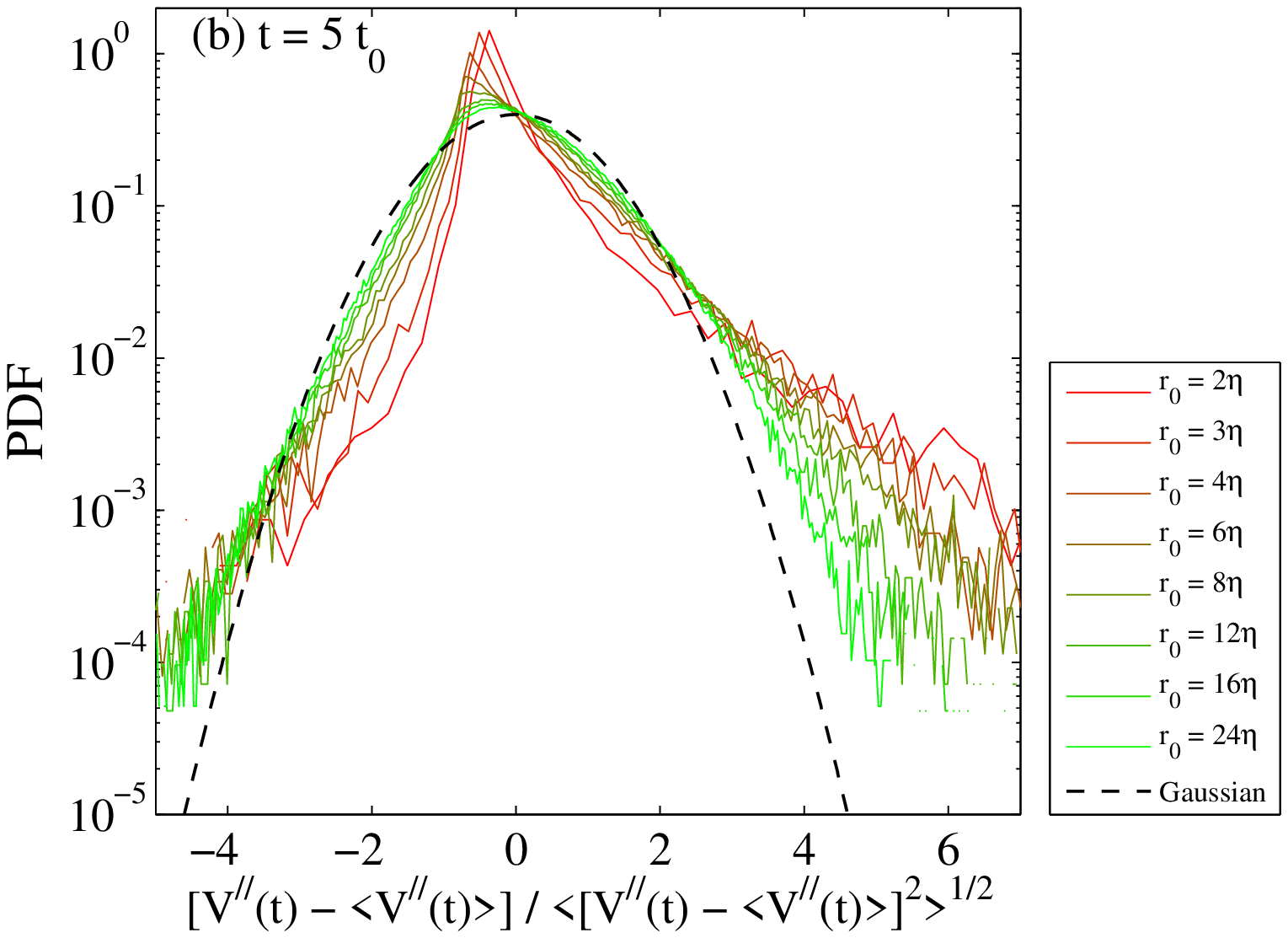}
  \caption{\label{fig:histo_Vpar} Centered and normalized probability
    density functions of the longitudinal velocity difference: (a) for
    $r_0=12\,\eta$ and various times and (b) for various initial
    separations and $t=5\,t_0$. In both cases, the black dashed line
    shows a Gaussian distribution.}
\end{figure}

All the considerations on the dependence of $V^\parallel$ upon the
initial separation are also visible in the probability distribution of
the longitudinal velocity.  Figure~\ref{fig:histo_Vpar}~(a) shows the
centered PDF of $V^\parallel$ normalized to unit variance for various
times and $r_0=12\eta$. The data clearly show that at times later than
$0.01\,t_0$, there is a change in the sign of the skewness. Also, one
sees that the time dependence of the skewness and of the flatness
comes from the right tails associated to large velocity differences,
supporting the arguments discussed above. The left tails, which
corresponds to approaching pairs, seem on the contrary to
collapse. The right tail has a very rich behavior. It starts with
broadening at times $t<t_0$, in agreement with the increase of
flatness. For $t>t_0$, it then decreases and possibly goes back
asymptotically to its initial form. If this was true, it would imply
that the distribution of velocity differences keep in memory the
initial separation at any later time.

This is also evidenced from Fig.~\ref{fig:histo_Vpar}~(b), which shows
the same PDFs for various separations and a time $t=5\,t_0$
fixed. Again we observe a rather good collapse of the tails associated
to negative longitudinal velocity differences, but the right tails
display very strong dependence on the initial separation. Clearly, the
behavior of this tail is a stretched exponential for $r_0\lesssim
8\eta$ and is faster than exponential for larger initial
separations. The actual level of statistics do not allow us to relate
systematically this behavior to that of the initial velocity
difference distribution.

\begin{figure}[t]
  \includegraphics[width=\textwidth]{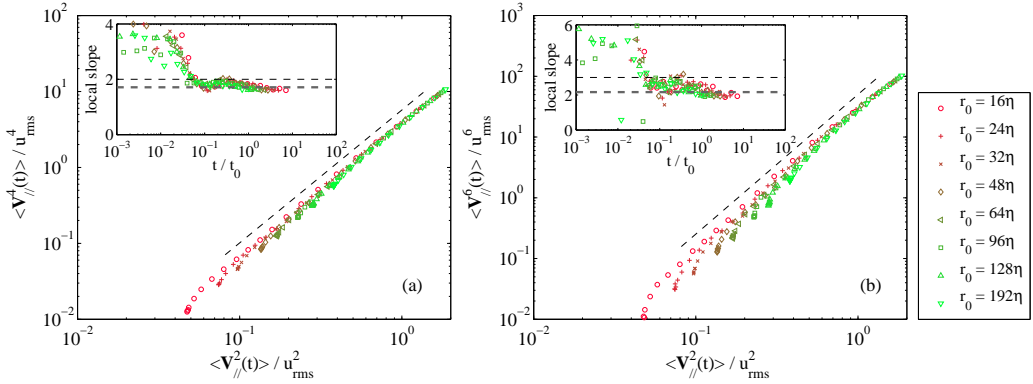}
  \caption{\label{fig:Vpar_intermittency} Fourth-order (a) and
    sixth-order (b) moments of the longitudinal velocity difference as
    a function of its second-order moment for various times and
    initial separations.  The two dashed lines correspond to a scaling
    compatible with that of Lagrangian structure functions proposed in
    \cite{biferale-etal:2004}, namely
    $\zeta^\mathrm{L}_4/\zeta^\mathrm{L}_2 = 1.71$ and
    $\zeta^\mathrm{L}_6/\zeta^\mathrm{L}_2 = 2.16$. The insets show
    the logarithmic derivative $\mathrm{d}\log
    \langle[V^\parallel(t)]^p\rangle / \mathrm{d}\log
    \langle[V^\parallel(t)]^2\rangle$ for (a) $p=4$ and (b) $p=6$ as a
    function of $t/t_0$; there the bold dashed lines show the
    Lagrangian multifractal scaling and the thin lines what is
    expected from a self-similar behavior.}
\end{figure}
Finally another way to address the question of intermittency of the
velocity difference consists in finding how moments of its
longitudinal component depend on time. For that we follow, as in the
case of the moments of distances, an approach similar to that of
extended self-similarity. Figures~\ref{fig:Vpar_intermittency}~(a) and
(b) show the fourth and sixth-order moments of $V^\parallel(t)$ as a
function of its second-order moment. As evidenced in the insets, they
display an anomalous behavior that differs from simple
scaling. However, the collapse for various $r_0$ is much less evident
than for the moments of the distance, except perhaps at sufficiently
large times. One can there guess a power-law dependence of $\langle
[V^\parallel(t)]^{p}\rangle$ as a function of $\langle
[V^\parallel(t)]^{2}\rangle$. Surprisingly the power is compatible
with the scaling exponent of the Lagrangian structure functions that
were obtained in \cite{biferale-etal:2004} by relating velocity
increment along trajectories to She--L\'ev\^eque multifractal spectrum
for the Eulerian field. The two dashed lines in
Fig.~\ref{fig:Vpar_intermittency}~(a) and (b) correspond to the
predicted values $\zeta^\mathrm{L}_4/\zeta^\mathrm{L}_2 = 1.71$ and
$\zeta^\mathrm{L}_6/\zeta^\mathrm{L}_2 = 2.16$. Confirming further
this match would require much better statistics.

%%%%%%%%%%%%%%%%%%%%%%%%%%%%%%%%%%%%%%%%%%%%%%%%
\subsection{Stationary distribution of rescaled velocity differences}

\begin{figure}[h]
  \includegraphics[width=0.5\textwidth]{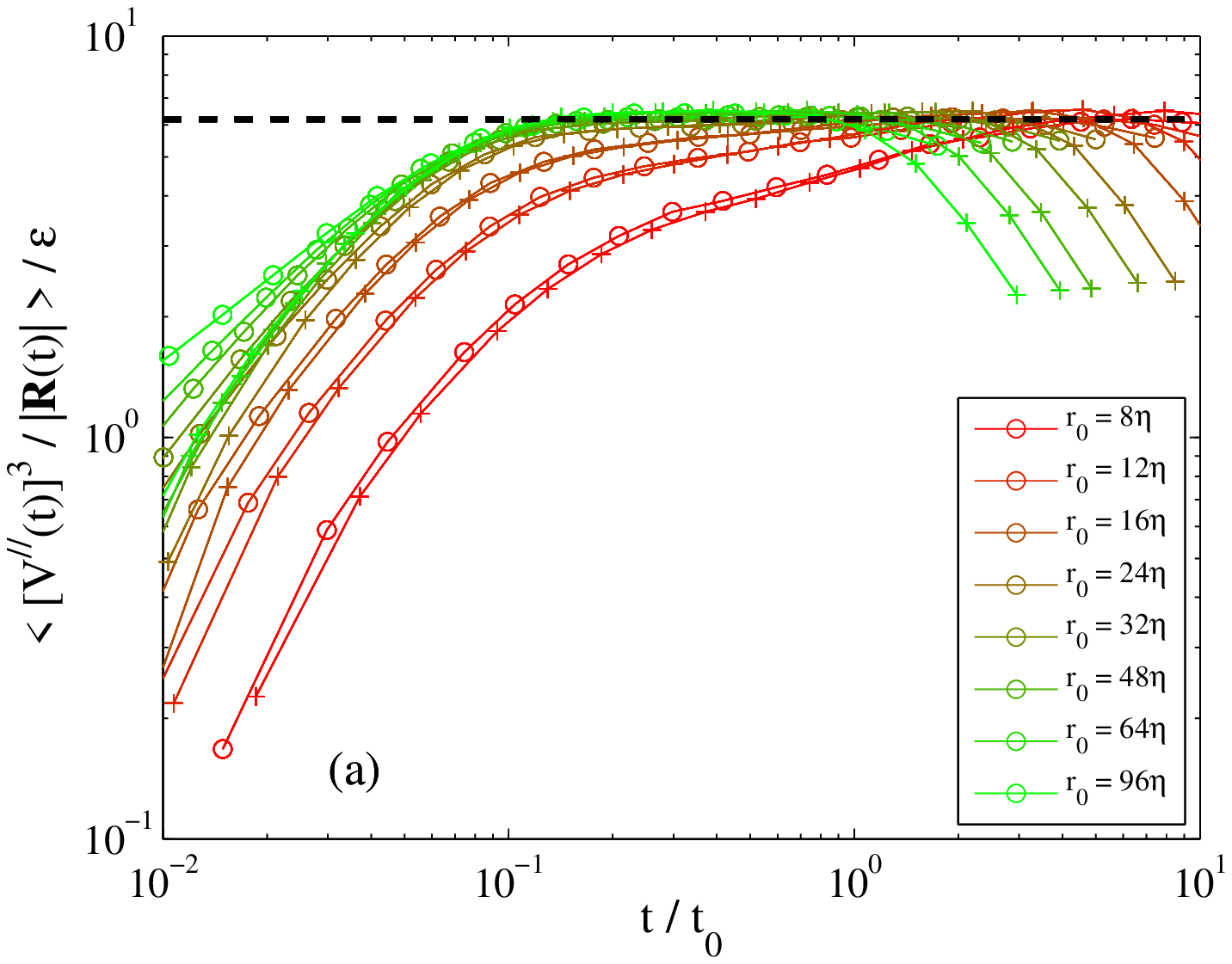}
  \hfill
  \includegraphics[width=0.5\textwidth]{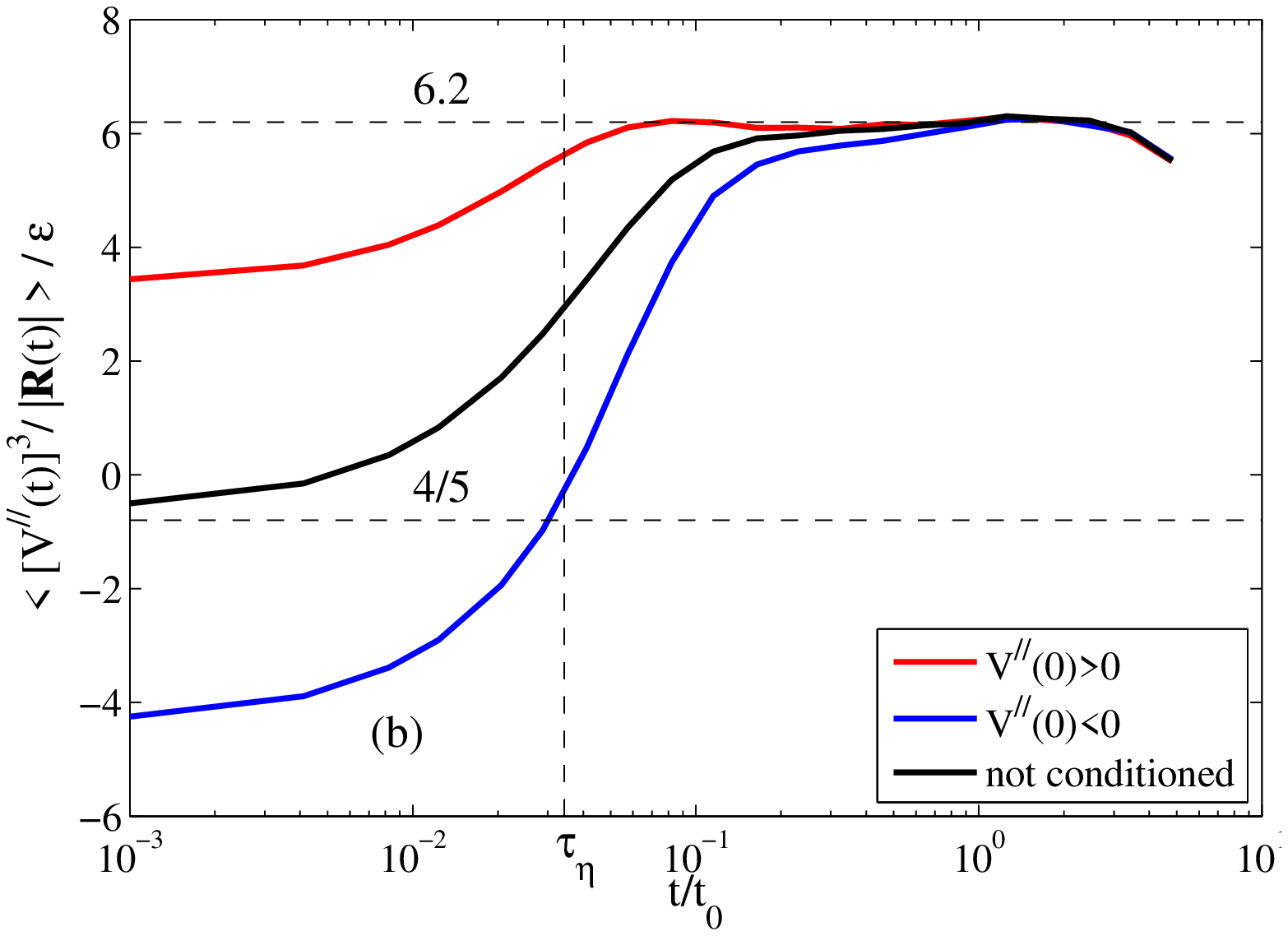}
  \caption{\label{fig:locdiss_fntime_r0_24eta} (a) Time evolution of
    the mixed moment $\langle [V^\parallel(t)]^3/|\bR(t)|\rangle$ for
    different initial separations and the two values of the Reynolds
    number: $R_\lambda = 730$ ($\circ$) and $R_\lambda = 460$
    ($+$). The black dashed line shows the asymptotic value
    $6.2\,\epsilon$. (b) Same for $r_0=24\,\eta$ and $R_\lambda = 730$
    in black and for the same moment but conditioned on positive (red
    curve) and negative (blue curve) values of the initial
    longitudinal velocity difference $V^\parallel(0)$.}
\end{figure}

As we have seen previously, the velocity difference between tracers
displays very intermittent features and, as a consequence, does not
converge to a behavior with temporal self-similarity, or does it only
very slowly. The situation is very different when interested in mixed
statistics between distances and longitudinal velocity differences. As
seen in~\cite{bitane-etal:2012}, the moment $\langle
[V^\parallel(t)]^3/|\bR(t)|\rangle$, which is initially negative and
equal to $-(4/5)\epsilon$, tends very quickly to a positive
constant\,---\,see Fig.~\ref{fig:locdiss_fntime_r0_24eta} (a).
The decrease at very large times comes from the contamination of the
statistics by pairs that have reached a distance of the order of the
integral scale.  The asymptotic value $\approx 6.2\,\epsilon$ seems to
depend only weakly on the Reynolds number. The collapse of the curves
associated to different Reynolds numbers and, for $r_0\gg\eta$, to
various initial separations indicate that the time of convergence is
$\propto t_0$.  Figure~\ref{fig:locdiss_fntime_r0_24eta} (b) shows the
same moment but conditioned on the sign of the initial longitudinal
velocity difference. One observes that for initially separating pairs
(red curve), the convergence to the asymptotic value is on a time of
the order of $\tau_\eta$.  Conversely for tracers that initially
approach each other (blue curve), the convergence is less fast. We
have seen in Sec.~\ref{subsec:geom_long} that such pairs first attain
their minimal distance at $t\approx t_\ast \approx 0.01\,t_0$. Then,
at that time, $\langle [V^\parallel(t)]^3/|\bR(t)| \,|\,
V^\parallel(0)<0\rangle$ changes sign and the convergence to $\approx
6.2\epsilon$ occurs only later. Such initially approaching pairs are
leading the average so that the overall convergence is on timescales
of the order of $t_\ast$. The mixed moment $\langle
[V^\parallel]^3/|\bR| \rangle$, which is a kind of ``local
dissipation'' along pairs of trajectories, is thus conserved by the
Lagrangian flow.

\begin{figure}[b]
  \includegraphics[width=\textwidth]{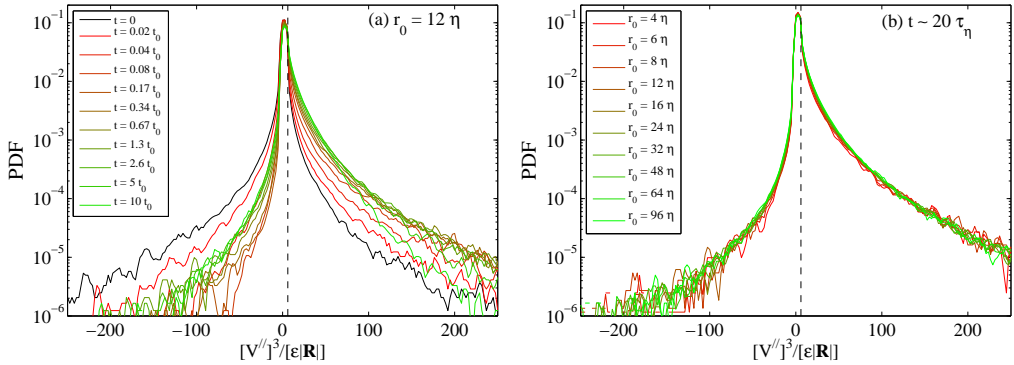}
  \caption{\label{fig:pdf_loc_diss} (a) PDF of the ``local
    dissipation'' $[V^\parallel(t)]^3/|\bR(t)|$ for $r_0=24\,\eta$ and
    various times (as labeled). (b) Same for various initial
    separations and a time $t=20\,\tau_{\eta}$ fixed. $t/t_0$ goes
    here from 0.2 to 5 so that, in all cases, the mixed quantity
    $[V^\parallel]^3/|\bR|$ has reached its asymptotic
    regime. In both figures the vertical dashed lines show the
    position of the average value $\approx 6.2\epsilon$.}
\end{figure}
Actually, it is not only the average of the ``local dissipation'' that
converges to a constant but its full distribution seems to attain a
stationary regime on times of the order of
$t_\ast$. Figure~\ref{fig:pdf_loc_diss} (a) shows for a given initial
separation, the convergence at large times of the PDF of
$[V^\parallel(t)]^3/|\bR(t)|$. One observes that the right and left
tails converge on different timescales. The tail associated to large
positive values (separating pairs) occurs on timescales of the order
of $\tau_\eta$ (which is in this case $\approx0.07\,t_0$), while that
for negative values (approaching pairs) converges slower. For the
largest time ($t\approx 5\,t_0$) one observes that deviations to the
asymptotic distribution occur again at very large positive
values. This is due to a contamination of such events by the large
scales of the turbulent flow. This decrease is in agreement with the
observed departure in Fig.~\ref{fig:locdiss_fntime_r0_24eta} of the
average from its asymptotic value at large
times. Figure~\ref{fig:pdf_loc_diss} (b) shows the PDFs of
$[V^\parallel]^3/|\bR|$ for different initial separations and at a
fixed time sufficiently large to be ensured that all distributions
have attained their asymptotic regime. One observes a robust collapse,
much more pronounced than for both the distribution of separations and
that of velocity differences. Note that in Fig.~\ref{fig:pdf_loc_diss}
the distributions are raw and were not rescaled by any moment of
$[V^\parallel]^3/|\bR|$. The asymptotic PDF is peaked around zero
(rather than its mean value) and displays fat tails that, according to
our data, are $\propto\exp(-C\,|V^\parallel|/|\bR|^{1/3})$ on both
sides.

To our knowledge, such a fast and manifest convergence of these mixed
statistics has never been reported before. From a phenomenological
viewpoint, one expects $V^\parallel \propto t \propto |\bR|^{1/3}$ in
the explosive Richardson--Obukhov regime, so that the ``local
dissipation'' $[V^\parallel]^3/|\bR|$ should become constant at
sufficiently large times. However we have observed here that the
convergence of this quantity to its asymptotic value occurs much
faster and in a much more definite manner than the convergence of the
statistics of $|\bR|$ and $|\bV|$ to their respective asymptotic
forms. This indicates that the statistical stationarity of the ``local
dissipation'' is more likely to be a cause rather than a consequence
of Richardson--Obukhov explosive separation. At the moment we
unfortunately lack a clear understanding of the physical mechanisms
that are responsible for the observed behavior of
$[V^\parallel]^3/|\bR|$ and that could shed light on its relation to
explosive separation.

%%%%%%%%%%%%%%%%%%%%%%%%%%%%%%%%%%%%%%%%%%%%%%%%
\section{Extreme events in separation statistics}
\label{sec:extreme}
\begin{figure}[h]
  \centering
  \includegraphics[width=\textwidth]{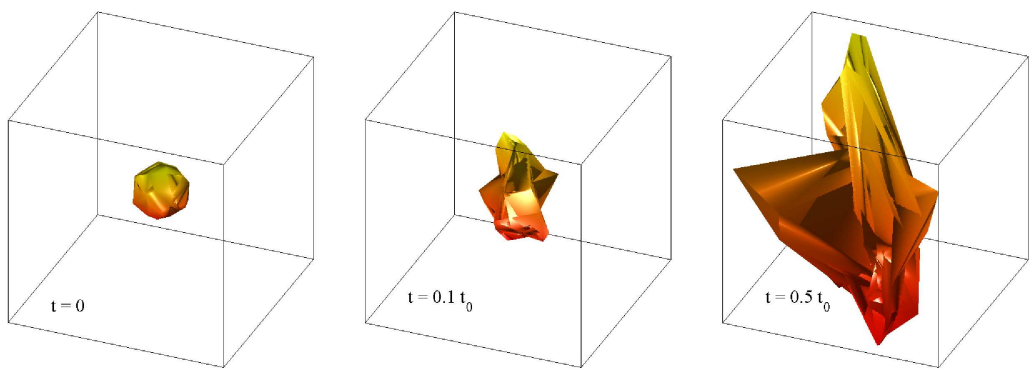}
  \caption{\label{fig:evol_sphere} Quasi-Lagrangian evolution of the
    set of tracers that are initially at a distance $r_0 = 24\eta$ of
    a reference trajectory.}
\end{figure}

In this section we turn back to the statistics of the distance
$|\bR(t)|$ between tracers. Our goal is to explain the mechanisms
leading to very large or very small values of this distance in the
light of the various observations made in previous
sections. Considering the relative dispersion of tracers conditioned
on their initial distance $r_0$ can be geometrically interpreted in
terms of the time evolution of an initially spherical surface of
radius $r_0$ that is centered on a reference trajectory. The transport
of this surface by a turbulent flow is generally very
complex. Incompressibility implies that the volume of the sphere is
conserved but the velocity field roughness will be responsible for
strong distorsions of its surface. This is represented in
Fig.~\ref{fig:evol_sphere} that shows for $R_\lambda = 460$ at three
various times the shape defined by the instantaneous position of 60
trajectories that are all initially at a distance $24\eta$ from a
reference tracer.
We can qualitatively deduce from these snapshots that the large
excursions of the inter-trajectory distance go together with strong
pinches of the surface. We also see that the intense stretchings are
occurring in a time-correlated manner: the surface angles that are
visible at late time have formed at very early stages. As we will now
see, the most-separated pairs have been so for long times and carry a
reminiscent dependence on the initial separation.

%%%%%%%%%%%%%%%%%%%%%%%%%%%%%%%%%%%%%%%%%%%%%%%%
\subsection{Memory in large-distance statistics}
\label{subsec:large-dist}

We first consider the events related to distances that are much larger
than their average. It was argued in Sec.~\ref{subsec:higher-order}
that the large-value tail of the separation PDF is not well described
by Richardson's distribution. As seen in Fig.~\ref{fig:histo_R_4096}
for $t=5\,t_0$, the tail is indeed broader than
$\exp(-C\,|\bR|^{2/3})$ for $r_0\lesssim 8\eta$ and narrower
otherwise.  It seems to tend to a Gaussian when either $r_0$ or $t$
are sufficiently large. We have checked that these behaviors are not
due to a contamination by pairs that have reached the large scales of
the turbulent flow, as the considered distances are still well below
$L$, except maybe for the largest initial separation
$r_0=24\eta$. These observations suggest that for such extreme events,
there is a long-term memory of the initial separation.

\begin{figure}[b]
  \includegraphics[width=\textwidth]{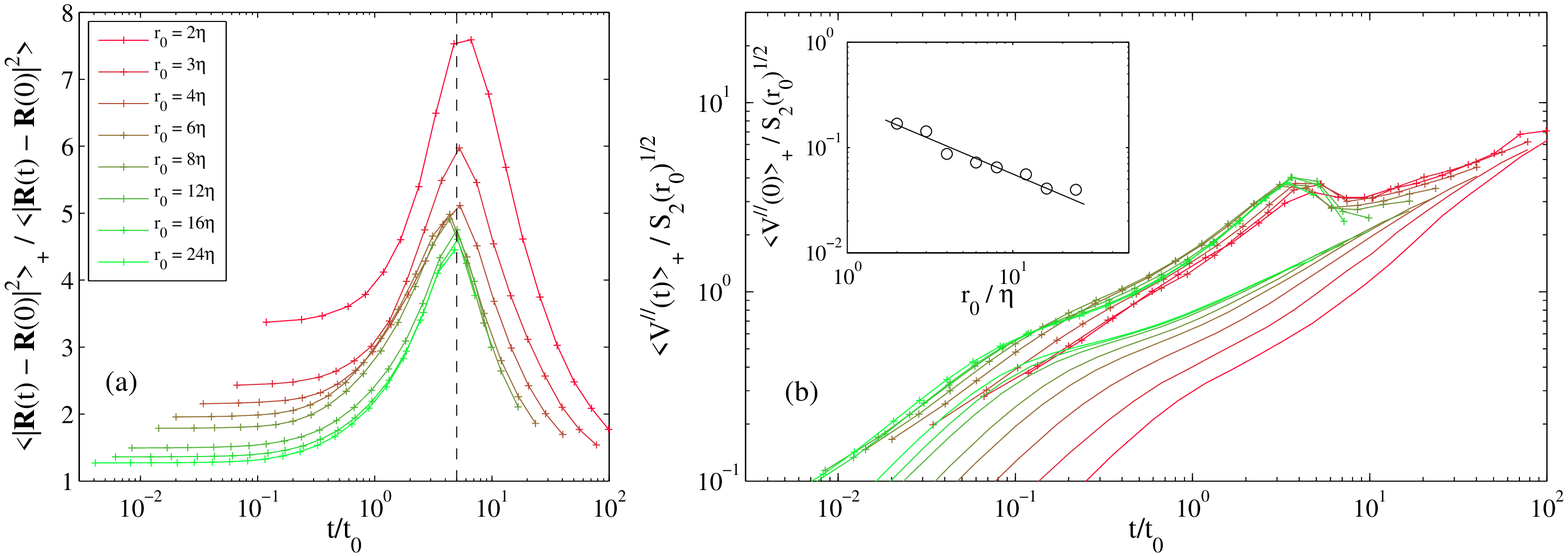}
  \caption{\label{fig:v_condr_fn_time} Conditional statistics over
    pairs that are at a distance at least twice the average at time
    $t_\mathrm{f}=5t_0$ (conditional average on $|\bR(5t_0)| \ge 2\,
    \langle |\bR(5t_0)| \rangle$ is denoted by
    $\langle\cdot\rangle_{+}$). (a) Relative increase in the mean
    squared change of separation $\langle|\bR(t)-\bR(0)|^2\rangle$ as
    a function of time. (b) Averaged longitudinal velocity difference
    conditioned on the separation at time $t_\mathrm{f}=5t_0$ ($+$)
    and without conditioning (solid line). Inset: initial value of the
    conditioned longitudinal velocity difference as a function of the
    initial separation; the solide line is $\propto r_0^{-2/3}$.}
\end{figure}

To qualify further the history of pairs that are well separated at
large times, we have performed the following analysis. Fixing a final
time $t_\mathrm{f}$ sufficiently large to have reached the explosive
Richardson--Obukhov regime, we have carried out statistics conditioned
on pairs that are far separated at $t=t_\mathrm{f}$, say such that
their distance is $|\bR(t_\mathrm{f})| \ge 2\, \langle
|\bR(t_\mathrm{f})| \rangle$. In order to not be contaminated by
finite inertial subrange effects, we have restricted this analysis to
$R_\lambda=730$ and to initial separations $2\,\eta\le r_0\le
24\,\eta$, and we have chosen the largest compatible value of
$t_\mathrm{f}$, namely $t_\mathrm{f} = 5\,t_0$.  Let us denote by
$\langle \cdot\rangle_{+}$ the resulting conditional ensemble average,
i.e.\ $\langle \cdot\rangle_{+}= \left\langle \cdot \,\mid\,\{
  |\bR(t_\mathrm{f})| \ge \langle |\bR(t_\mathrm{f})|
  \rangle\}\right\rangle$. Figure~\ref{fig:v_condr_fn_time}~(a)
represents the relative increase
$\langle|\bR(t)\!-\!\bR(0)|^2\rangle_+ /
\langle|\bR(t)\!-\!\bR(0)|^2\rangle$ of the mean-squared
displacement. The various curves, which are associated to different
initial separations $r_0$, have a maximum at $t = t_\mathrm{f} =
5\,t_0$ represented by a dashed line. The value of this
maximum is becoming larger when $r_0$ decreases and approaches the
dissipative scales. This is a signature of the dependence of the
large-positive value tail upon $r_0$. On the right-hand side of the
maximum, the various curves tend back to $1$. This indicates that
separations that are large at a given intermediate time relax in
average to an ordinary behavior at later times. This can be regarded
as a consequence of the weakening of large-distance probability tails
as a function of time.
The situation is rather different when interested in the left-hand
side of the maximum. The curves do not converge to $1$, and thus to an
average behavior when $t\to0$. This means that pairs that are well
separated at a given time are likely to have been so at any previous
times. This asymmetry when going forward or backward in time could
already be grasped in Fig.~\ref{fig:evol_sphere}. The trajectories
that are far away from the reference tracer at the latest time (as,
e.g., those defining the left and top corners) are visibly also well
separated at the intermediate time. Conversely the right corner at the
intermediate time have later stopped separating faster than the
average.

Figure~\ref{fig:v_condr_fn_time} (b) shows the time evolution of the
averaged longitudinal velocity difference $\langle
V^\parallel(t)\rangle_+$ conditioned on having a separation twice the
average at time $t_\mathrm{f} = 5\,t_0$. A first remark that cannot be
seen from the log-log plot is that the initial value of this average
is strictly positive. The inset represents the variation of $\langle
V^\parallel(0)\rangle_+$ as a function of the initial separation. This
implies that the pairs that are well separated at a late time are
preferentially separating from the very beginning. The initial
longitudinal velocity fluctuation that is necessary for the pairs to
be far apart at time $t_\mathrm{f}$ becomes weaker when $r_0$
increases.  Data suggest that $\langle V^\parallel(0)\rangle_+ /
[S_2(r_0)]^{1/2} \propto r_0^{-2/3}$ as seen in the inset. This
initial separation makes the selected pairs reach an almost diffusive
regime at time much shorter than the end of the average ballistic
regime. For large-enough initial separations we indeed have $\langle
V^\parallel(t)\rangle_+ \sim t^{1/2}$ for $0.1\,t_0\lesssim t \lesssim
t_0$. The pair distance encountered a final acceleration right before
the time of conditioning $t_\mathrm{f}$. After that, the longitudinal velocity
difference relaxes slowly to the average regime. Again here, as in the
case of the average separation, we observe that the imposing of having
a large distance at a large time selects pair histories. The main
contribution to statistics is indeed given by couples encountering an
initially violent separation and quickly reaching larger scales. Such
pairs converge rapidly to an explosive regime and continue to separate
faster than the average for a long time. This is the snowball effect
we have mentioned before.

\begin{figure}[h]
  \centering
  \includegraphics[width=0.6\textwidth]{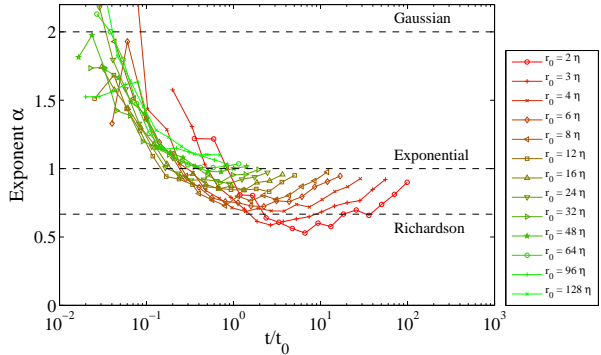}
  \caption{\label{fig:exp_rigthtail} Exponent $\alpha$ of the right
    tail of the PDF of distances $|\bR(t)|$ as a function of time and
    for various initial separations. The exponent was obtained by
    fitting $-\log p(|\bR|)$ to a power-law for
    $\langle|\bR(t)|\rangle < |\bR(t)| < 0.5\,L$.}
\end{figure}
In order to give a more quantitative handle on the far tail of the
separation PDF, we have estimated and fitted its functional shape and
studied its variations.  For that, following the observations made in
Sec.~\ref{subsec:higher-order}, we have assumed that $p(|\bR|) \propto
\exp (-C\,|\bR|^\alpha)$ and measured how the exponent $\alpha$
depends on both time and initial separation.
Figure~\ref{fig:exp_rigthtail} shows the time behavior of the exponent
$\alpha$ for various values of the initial separation. On the figure,
the three dashed horizontal lines represent the tails of a Gaussian
($\alpha =2$), of an exponential ($\alpha=1$) and of Richardson's
distribution ($\alpha=2/3$). At time $t=0$, the distribution is peaked
around $r_0$ and compactly supported, so that $\alpha=\infty$. The
exponent then decreases, crosses $\alpha=1$ (so that the distribution
becomes a stretched exponential) for $t\approx 0.1\,t_0$, and reaches
a minimal value that depends on the initial separation $r_0$. For the
smallest $r_0$, this minimal value is below Richardson's prediction,
as previously noticed. For larger separations, this minimum increases
and when $r_0\gg\eta$ the curves seem to saturate to the value $\alpha
= 1$. Also, we cannot exclude that all curves converge to the
exponential value when $t\to\infty$. The increase at the last stage
can hardly be blamed on an integral-scale effect. We have indeed
excluded here all pairs that are separated by a distance larger than
$L/2$.

%%%%%%%%%%%%%%%%%%%%%%%%%%%%%%%%%%%%%%%%%%%%%%%%
\subsection{``Fractal distribution'' at small distances}
\label{subsec:fractal}
%%%%%%%%%%%%%%%%%%%%%%%%%%%%%%%%%%%%%%%%%%%%%%%%

\begin{figure}[h]
  \includegraphics[width=0.5\textwidth]{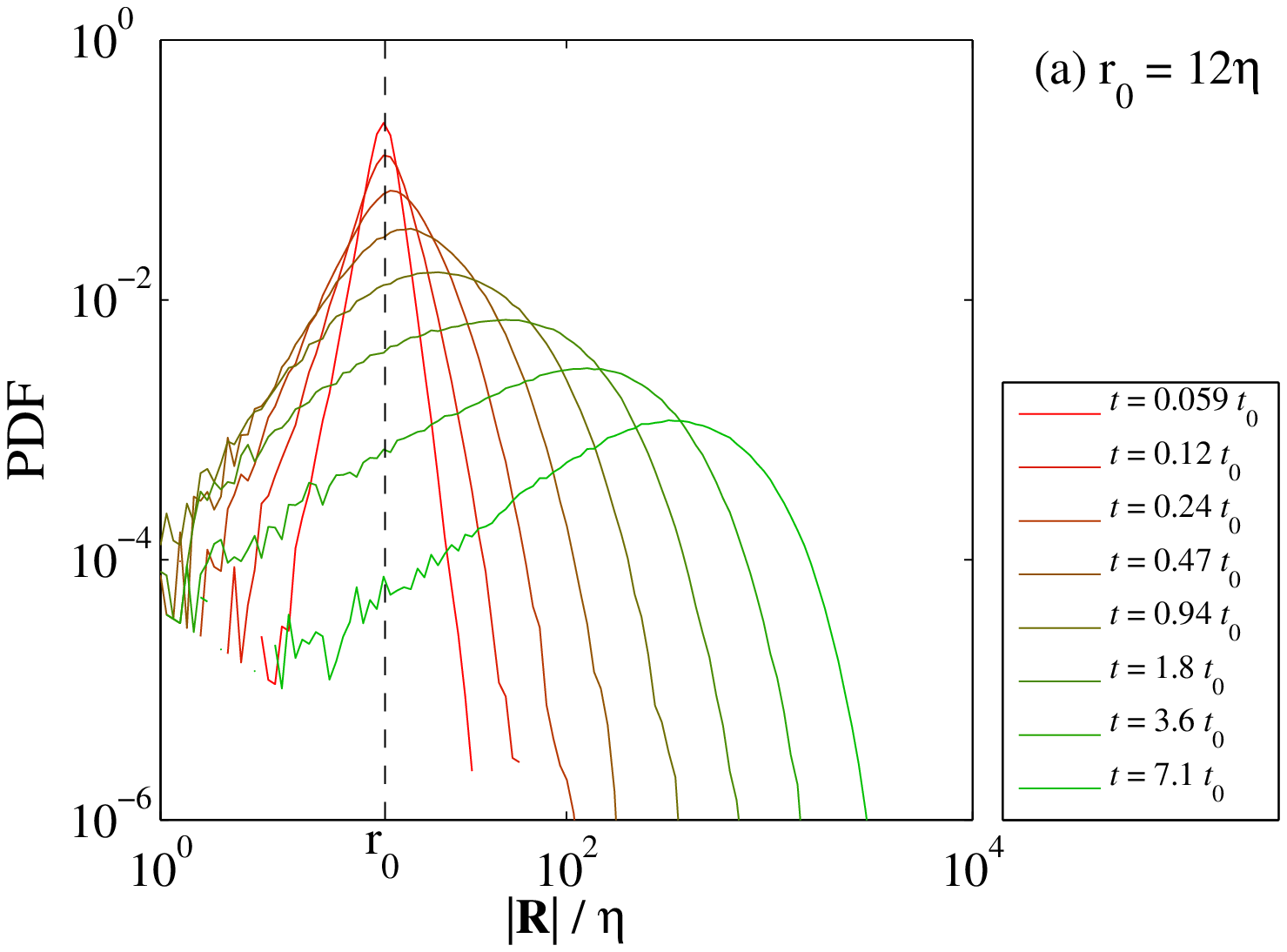}\hfill
  \includegraphics[width=0.5\textwidth]{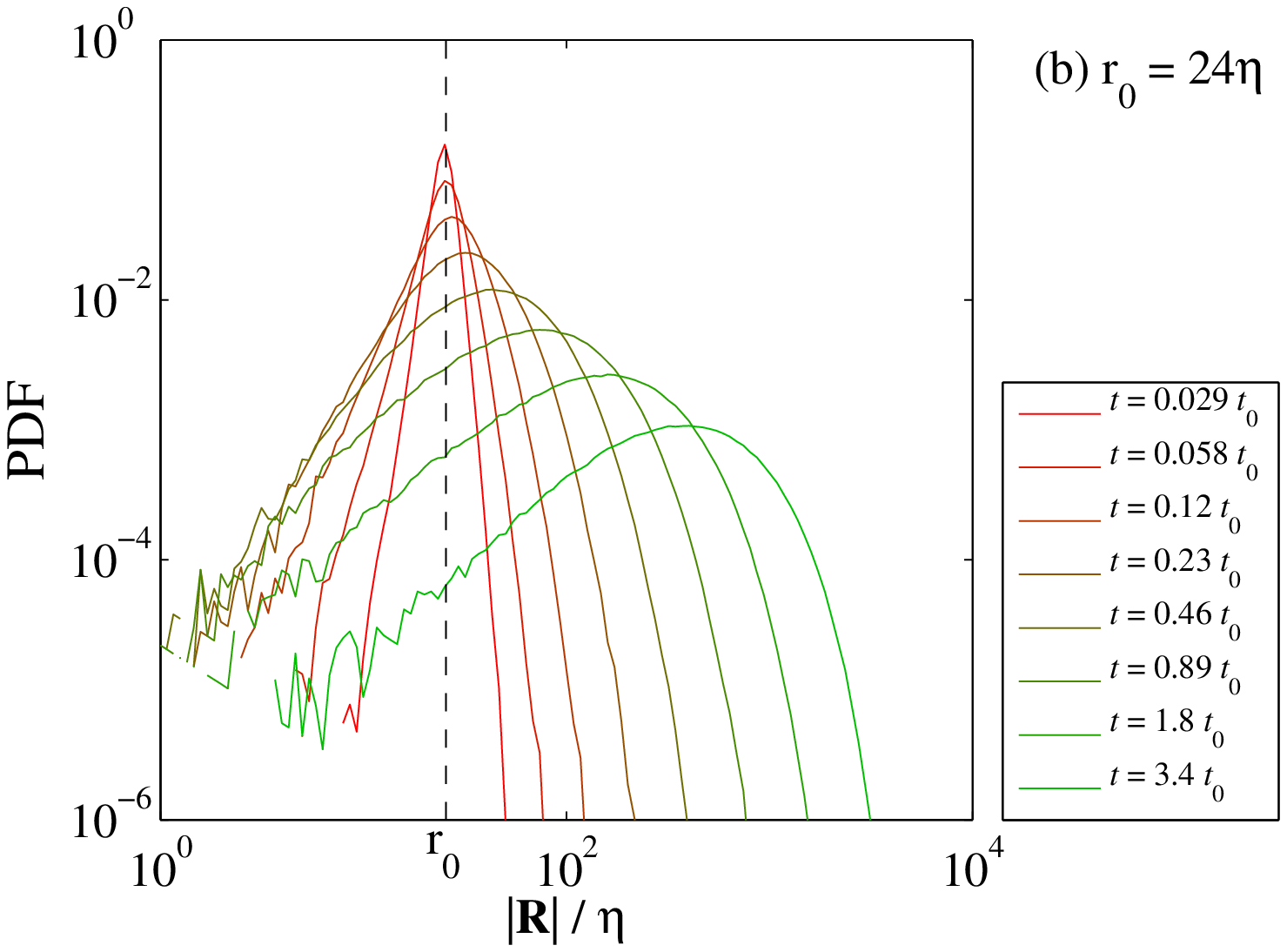}
  \caption{\label{fig:histo_r_notscaled} Unscaled PDFs of the absolute
    separation between tracers for various times and the initial
    separations (a) $r_0=12\,\eta$ and (b) $r_0=24\,\eta$.}
\end{figure}
Much more violent and intermittent events take place for particle
pairs that separate much less than the average. As already stressed
and observed in Fig.~\ref{fig:evol_sphere}, the large excursions of
inter-trajectory distances go together with strong pinches of
separations.  This is evidenced from Fig.~\ref{fig:histo_r_notscaled}
that represents the probability distribution of $|\bR|$ at various
times and for two inertial-range values of the initial separation. One
observes that at very short times, the PDF is peaked around
$r_0$. When time increases, its maximum, which roughly corresponds to
the value of $\langle|\bR(t)|\rangle$, shifts to larger values and the
distribution broadens simultaneously at large and small values. This
leads to the development for $r_0\ll |\bR| \ll \langle|\bR|\rangle$ of
an intermediate range of pairs whose separations lag behind the
average evolution. In this subrange, the PDF behaves as a power law
$p(|\bR|) \propto |\bR|^{\beta}$, where the exponent $\beta$ evolves
as a function of time and $r_0$. The power-law behavior is
substantiated when measuring the cumulative probability $P^<(r) =
\mathrm{Prob}[|\bR|<r] = \int_0^r p(r^\prime)\,\mathrm{d}r^\prime
\propto r^{\beta+1}$ of inter-trajectory distances. One expects from
Richardson arguments that $\beta=2$, so that $P^<(r) \propto
r^{3}$.

\begin{figure}[b]
  \centering
  \includegraphics[width=0.75\textwidth]{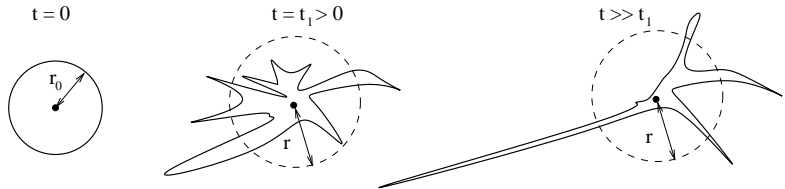}
  \caption{\label{fig:sketch_fractal} Two dimensional sketch of the
    time evolution of trajectories that are initially located at a
    distance $r_0$ of a reference tracer (shown as a black dot at the
    center of the circle) and then spread.}
\end{figure}
The exponent $\beta$ can be interpreted in terms of fractal
geometry. If all the trajectories were uniformly and independently
distributed in space, the fraction of pairs at a distance less than
$r$ would be $\propto r^{d}$, where $d\!=\!3$ is the space
dimension. Also one would expect this fraction to be $\propto r^{d-1}$
if the trajectories were all confined on a surface, and $\propto
r^{d-2}$ if they were on a curve. In general, the exponent $\beta+1 =
\lim_{r\to 0} [\log P^<(r)]/[\log r]$ measures the \emph{correlation
  dimension} of a fractal set. Richardson's $r^2$ behavior thus
corresponds to the idea that, at sufficiently long times, trajectories
forget about their initial separation and distribute homogeneously in
space.  As we have previously discussed, pair dispersion can be
geometrically interpreted as the average Lagrangian evolution of an
initially spherical surface of radius $r_0$ centered on a reference
trajectory. At time $t=0$, it is clear that $P^<(r)$ is a Heaviside
function centered on $r=r_0$, so that $\beta=\infty$. At later times
the stretching and pinching of the sphere leads to a non-trivial
behavior of $P^<(r)$ for $r\ll\langle|\bR|\rangle$. This behavior is
sketched for two dimensions in Fig.~\ref{fig:sketch_fractal}. The
exponent $\beta$ is thus measuring the fractality of the image of the
sphere by the Lagrangian flow. It relates to the fraction of this
forward-in-time image that remains within a distance $r$ of the
reference trajectory. At sufficiently large times (right-most panel of
Fig.~\ref{fig:sketch_fractal}), one expects the set of trajectories to
be mainly stretched by the flow eddies with the most relevant
correlation time, that is those of size $\langle|\bR(t)|\rangle$, and
the small scale fluctuations to become less and less important. Hence,
at very large times, the only pairs that are still at a distance $r\ll
\langle|\bR(t)|\rangle$ should be distributed on a surface (curve in
two dimensions), so that $\beta+1 \simeq 2$ and $p(r) \propto r$. This
approach contradicts the prediction of Richardson that, somehow,
postulates that the velocity difference correlation time does not
depend on scale (as it is assumed to be zero).

\begin{figure}[h]
  \includegraphics[width=0.5\textwidth]{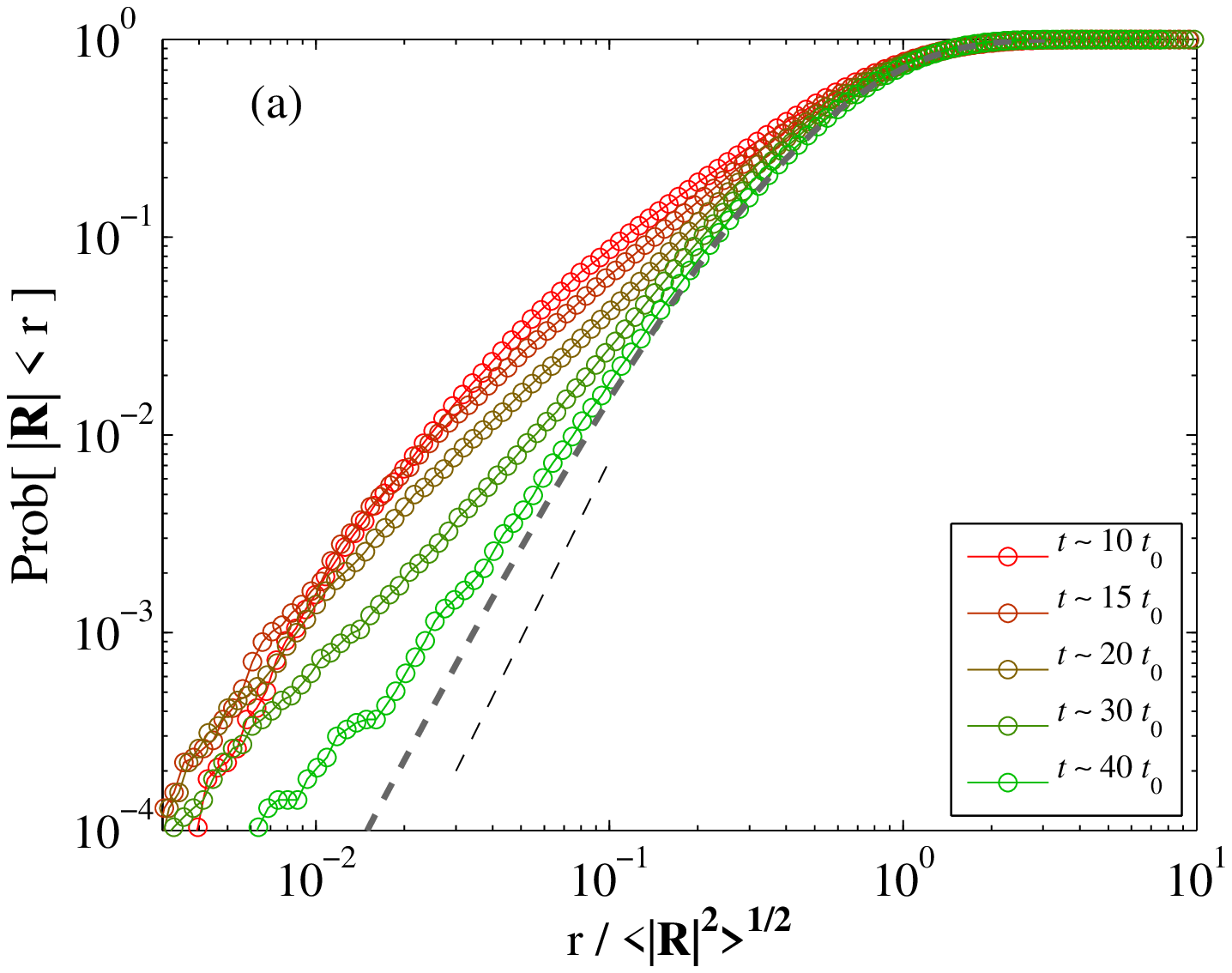}
  \includegraphics[width=0.5\textwidth]{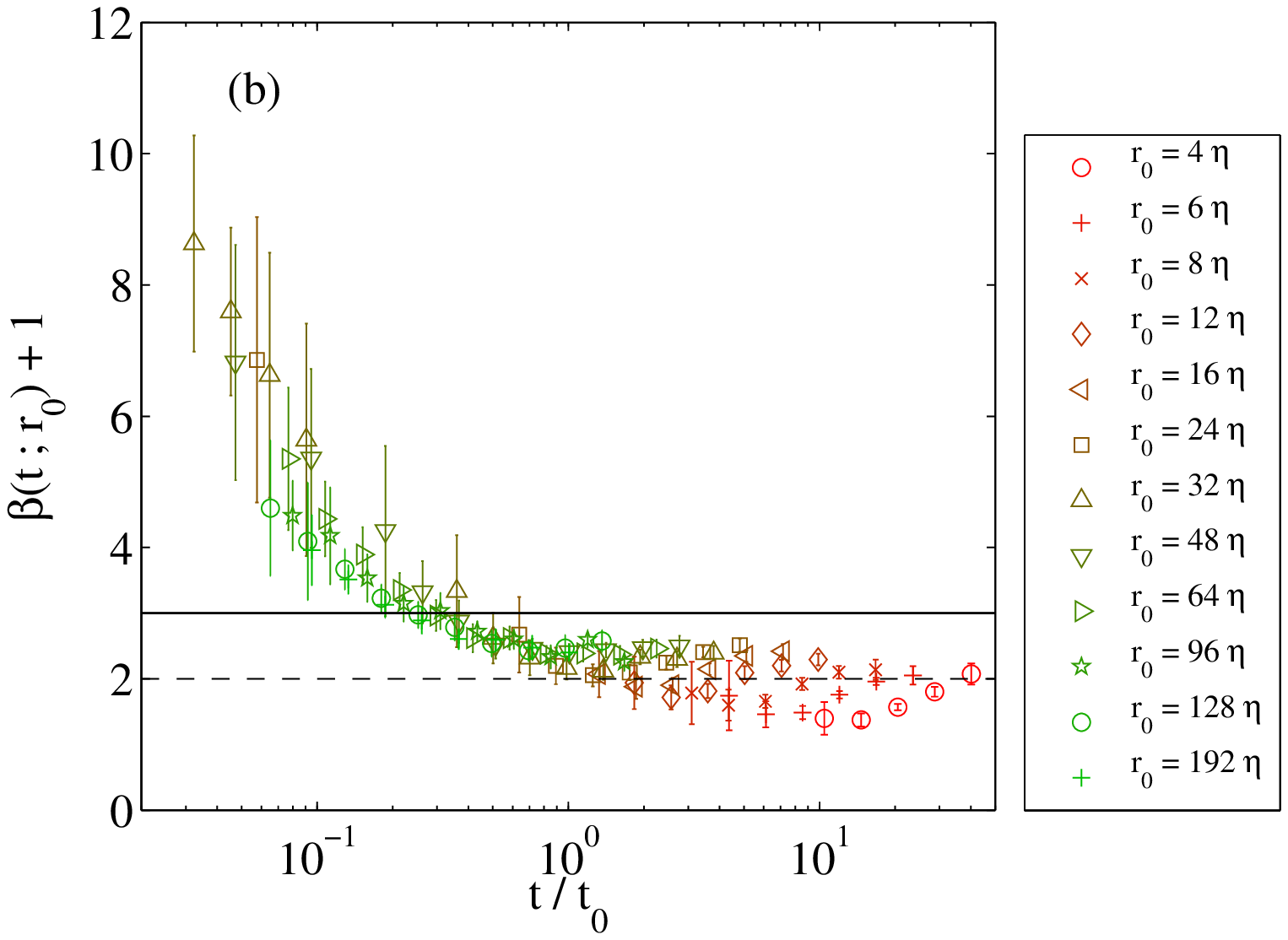}
  \caption{\label{fig:exponent} (a) Cumulative probability
    distribution of the distance $P^<(r) = \mathrm{Prob}[|\bR|<r]$ as
    a function of $r$ for $r_0=4\eta$ and at various times; the gray
    dashed curve is the cumulative probability associated to
    Richardson's distribution that displays a behavior $\propto r^3$ at
    small distances. (b) Time evolution of the left-tail power-law
    exponent $\beta$ for various values of the initial separation. The
    black solid line shows $\beta=2$ corresponding to Richardson's
    distribution and the dashed line stands for the value $\beta=1$
    that is argued in the text.}
\end{figure}
To shed light on such ideas, we have estimated the exponent $\beta$
for various $r_0$ and at different times.  For this purpose, we have
used the cumulative probability $P^<(r)$ (shown in
Fig.~\ref{fig:exponent}~(a) for $r_0=4\eta$) to obtain $\beta+1$
through a fit of its local slope. The resulting measurements are
displayed in Fig.~\ref{fig:exponent}~(b) as a function of time. Up to
the large error bars due to a lack of statistics, one observes that
the timescale $t_0$ is again relevant to describe both the initial
dynamics and the convergence to an asymptotic regime. At small times
$t\ll t_0$, the exponent $\beta$ takes large positive values, which,
as explained above, are due to the initial conditioning
$|\bR(t)|=r_0$.  At time $t\approx 0.3\,t_0$, the exponent becomes
smaller than 3. For $t\gg t_0$, it asymptotically approaches the value
$\beta=2$, so that data give evidence in favor of the argument exposed
above. Interestingly, for small initial separations $r_0\lesssim
16\eta$, the limit is reached from below and preceded by a minimum of
$\beta$. This means that for such intermediate times, the pinching (or
equivalently the stretching) of the sphere is so strong that the
intersection of the latter with spheres of radius $r$ defines a
fractal object that is more concentrated than a surface. Such a
temporary singular behavior is certainly related to the presence of
tails fatter than exponential at very large distances.

%%%%%%%%%%%%%%%%%%%%%%%%%%%%%%%%%%%%%%%%%%%%%%%%
\section{Concluding remarks}
\label{sec:conclusion}
%%%%%%%%%%%%%%%%%%%%%%%%%%%%%%%%%%%%%%%%%%%%%%%%

In this paper we have confirmed and extended previous results of
\cite{bitane-etal:2012} on the timescales of convergence of turbulent
pair dispersion to an asymptotic regime. We have seen that low and
medium-order moments of the particle separation approach an asymptotic
regime on times of the order of $t_0=S_2(r_0)/(2\epsilon)$, where
$S_2(r_0)$ denotes the (absolute value) structure function associated
to the initial separation $r_0$ and $\epsilon$ is the mean rate of
kinetic energy dissipation. This timescale has been shown to be
relevant to describe also the initial kinematic change of sign of the
longitudinal velocity difference $V^\parallel$ and the tails of the
distance distribution at small values. However, we have seen that
$t_0$ is not relevant to describe the convergence of the velocity
difference statistics to an asymptotic regime. We have observed that
up to the largest time, the skewness, the flatness, and more generally
the shape of the velocity difference distribution still depend on the
initial separation, even when time is rescaled by $t_0$.  This leads
to a behavior that is by far more intermittent than that of
separations. Also we obtained evidence that the far tail distribution
of separation is also keeping a nontrivial memory on $r_0$ up to the
largest times. Throughout this paper, we made an important use of
geometrical considerations to explain phenomenologically the
statistical events leading to extreme fluctuations. In particular, we
argued and obtained numerical evidence that, at sufficiently large
times, the probability distribution $p(r)$ of inter-tracer distances
is $\propto r$ for $r\ll\langle|\bR|\rangle$ and decays as
$\exp(-C\,r)$ when $r\gg\langle|\bR|\rangle$. These two observations
strongly contradicts Richardson's eddy diffusivity approach, which
seems nevertheless to give a good approximation of the core of the
distribution. Finally, we obtained a striking result concerning mixed
distance and velocity statistics. We indeed found that the
distribution of a ``local dissipation'', defined as
$[V^\parallel]^3/|\bR|$, attains relatively quickly a scaling regime
that is independent of both $r_0$ and time. This behavior, which, to
our knowledge, has never been observed before, gives very strong
constraints for the development and validation of stochastic Markovian
models for turbulent relative dispersion.

A central question that we raised concerns the physical mechanisms
leading to this fast convergence of the ``local dissipation'' to a
statistically stationary behavior. At the moment, an even
phenomenological explanation is still missing. We have understood that
kinematic considerations can be used to explain why $\langle
[V^\parallel]^3/|\bR|\rangle$, which is initially negative, becomes
quickly positive. This is due to the fact that initially approaching
trajectories will always eventually separate. However, there is no
clear idea of why this average should converge to a time-independent
value. Possible ideas could consist in finding other mixed moments
that are strictly (and not asymptotically) conserved by the two-point
turbulent Lagrangian flow. Such considerations are kept for future
work. Finally, another possible extension of the current study is to
apply the developed understanding of the geometry of relative
dispersion to more complicated turbulent transport situations
involving more than two trajectories.  This is for instance the case
for the forward-in-time dynamics of triangles or tetrahedrons studied
in \cite{pumir-etal:2000}. The quality of today numerical data would
be very useful to revisit such questions with an emphasize on extreme
events. Another situation that is much closer to applications is that
of the transport by turbulence of pollutant patches. We have seen in
the current study that relative dispersion relates to the distorsion
of a sphere by the Lagrangian flow. We related the small-scale
behavior of separation statistics to the fractal dimension of this
object. Such geometrical considerations could be generalized to
understand and quantify the large concentration fluctuations in the
dispersion of a pollutant spot.

\section*{Acknowledgments}
We are grateful to L.~Biferale, M.~Cencini, G.~Falkovich,
A.~Frishmann, G.~Krstulovic, A.~Lanotte, S.S.~Ray for useful
discussions and remarks.  Access to the IBM BlueGene/P computer JUGENE
at the FZ J\"ulich was made available through the XXL-project
HBO28. The research leading to these results has received funding from
DFG-FOR1048 and from the European Research Council under the European
Community's Seventh Framework Program (FP7/2007-2013, Grant Agreement
no. 240579).

\bibliographystyle{tJOT}

\end{document}